\documentclass[lettersize,journal]{IEEEtran}
\usepackage{amsmath,amsfonts}
\usepackage{algorithmic}
\usepackage{algorithm}
\usepackage{array}
\usepackage{textcomp}
\usepackage{stfloats}
\usepackage{url}
\usepackage{verbatim}
\usepackage{graphicx}
\usepackage{cite}
\usepackage{microtype}
\usepackage{amsthm}
\usepackage{amsmath,amsfonts,bm}

\def\eqref#1{equation~\ref{#1}}
\def\1{\bm{1}}

\def\mA{{\bm{A}}}
\def\mB{{\bm{B}}}

\def\mH{{\bm{H}}}
\def\mI{{\bm{I}}}

\def\mK{{\bm{K}}}

\def\mP{{\bm{P}}}
\def\mQ{{\bm{Q}}}

\def\mS{{\bm{S}}}

\def\mV{{\bm{V}}}
\def\mW{{\bm{W}}}
\def\mX{{\bm{X}}}

\def\mZ{{\bm{Z}}}

\DeclareMathAlphabet{\mathsfit}{\encodingdefault}{\sfdefault}{m}{sl}
\SetMathAlphabet{\mathsfit}{bold}{\encodingdefault}{\sfdefault}{bx}{n}

\newcommand{\Ls}{\mathcal{L}}
\newcommand{\R}{\mathbb{R}}

\theoremstyle{plain}
\newtheorem{theorem}{Theorem}[section]

\theoremstyle{definition}
\newtheorem{definition}[theorem]{Definition}

\theoremstyle{remark}

\usepackage[textsize=tiny]{todonotes}

\usepackage{bbm}
\usepackage{mathtools}
\usepackage{amsmath}
\usepackage{amsthm}
\usepackage{wrapfig}
\usepackage{url}
\usepackage{graphicx}
\usepackage{subcaption} % subfigure

\usepackage{threeparttable} 
\usepackage{algorithm}
\usepackage{algorithmic}
\usepackage{colortbl}

\usepackage{authblk}
\begin{document}

\title{TokenMark: A Modality-Agnostic Watermark \\ for Pre-trained Transformers}

% \author{
%   % IEEE Publication Technology,~\IEEEmembership{Staff,~IEEE,}
%         % <-this % stops a space
% % \thanks{This paper was produced by the IEEE Publication Technology Group. They are in Piscataway, NJ.}% <-this % stops a space
% % \thanks{Manuscript received April 19, 2021; revised August 16, 2021.}

% }

\author[1]{Hengyuan Xu}
\author[1]{Liyao Xiang\thanks{Corresponding author: xiangliyao08@sjtu.edu.cn.}}
\author[1]{Borui Yang}
\author[2]{Xingjun Ma}
\author[1]{Siheng Chen}
\author[3]{Baochun Li}
\affil[1]{Shanghai Jiao Tong University}
\affil[2]{Fudan University}
\affil[3]{University of Toronto}

% % The paper headers
% \markboth{Journal of \LaTeX\ Class Files,~Vol.~14, No.~8, August~2021}%
% {Shell \MakeLowercase{\textit{et al.}}: A Sample Article Using IEEEtran.cls for IEEE Journals}

% \IEEEpubid{0000--0000/00\$00.00~\copyright~2021 IEEE}
% Remember, if you use this you must call \IEEEpubidadjcol in the second
% column for its text to clear the IEEEpubid mark.

\maketitle

\begin{abstract}
    Watermarking is a critical tool for model ownership verification. However, existing watermarking techniques are often designed for specific data modalities and downstream tasks, without considering the inherent architectural properties of the model. This lack of generality and robustness underscores the need for a more versatile watermarking approach. In this work, we investigate the properties of Transformer models and propose \textbf{TokenMark}, a modality-agnostic, robust watermarking system for pre-trained models, leveraging the permutation equivariance property. TokenMark embeds the watermark by fine-tuning the pre-trained model on a set of specifically permuted data samples, resulting in a watermarked model that contains two distinct sets of weights—one for normal functionality and the other for watermark extraction, the latter triggered only by permuted inputs. Extensive experiments on state-of-the-art pre-trained models demonstrate that TokenMark significantly improves the robustness, efficiency, and universality of model watermarking, highlighting its potential as a unified watermarking solution. 
\end{abstract}

\begin{IEEEkeywords}
Model watermarking, Transformer, intellectual property protection, permutation equivariance.
\end{IEEEkeywords}

% \input{sections/introduction.tex}

% \cite{steiner2021train} % dummy, delete this line later
\section{Introduction}
\label{sec:introduction}

\begin{figure*}[t]
    \centering
    \includegraphics[width=0.88\linewidth]{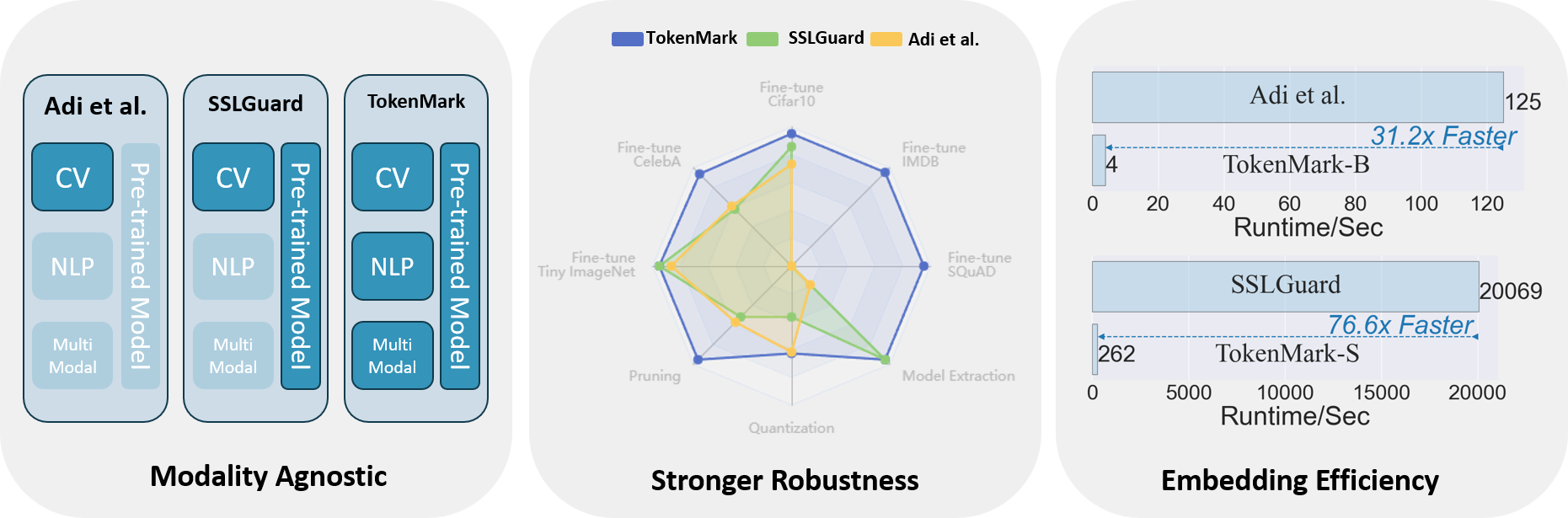}
    \caption{TokenMark is a modality-agnostic, robust, and lightweighted watermark for pre-trained models. It has a wider application range, and could serve as a universal plugin to replace the trigger in backdoor-based watermarking systems to enhance robustness against various removal attacks. 
    	}
    \label{fig:thumbnail}
\end{figure*}

\IEEEPARstart{T}{he} last few years have witnessed the revolutionary success of deep learning and pre-trained large models. High-performance large backbone models are usually trained with massive data, expert knowledge, and abundant computational resources, and thus are considered as proprietary assets. However, such assets are constantly facing the threats of being stolen, redistributed, or abused by adversaries. In that case, it is critical for the owner to provide convincing evidence to claim the model ownership. Watermarking serves as a promising tool for intellectual property (IP) protection of deep learning models \cite{lukas2022sok,jia2021entangled,abdelnabi2021awt,cong2022sslguard}. 

The watermarking schemes for data of different modalities typically vary significantly. For computer vision (CV) models, it is common to embed trigger samples as backdoors to the model, and later verify the watermark by examining if the model produces the targeted outputs on triggered samples \cite{adi2018turning,zhang2018protecting,jia2021entangled,lukas2022sok,cong2022sslguard}. For natural language processing (NLP) models, the watermark is embodied by the output distributions following some carefully-designed prompts \cite{kirchenbauer2023watermark,abdelnabi2021awt}. It is obvious that these watermarking schemes are tightly bond with the downstreaming tasks, which may not be suitable for pre-trained models fusing multiple modalities. For example, Transformers \cite{vaswani2017attention} are prevalent in CV, NLP, and multi-modal tasks, and thus any data-type-specific watermarking scheme cannot reflect the inherent nature of the model.

\IEEEpubidadjcol

\begin{figure}
	\centering
	\includegraphics[width=0.6\linewidth]{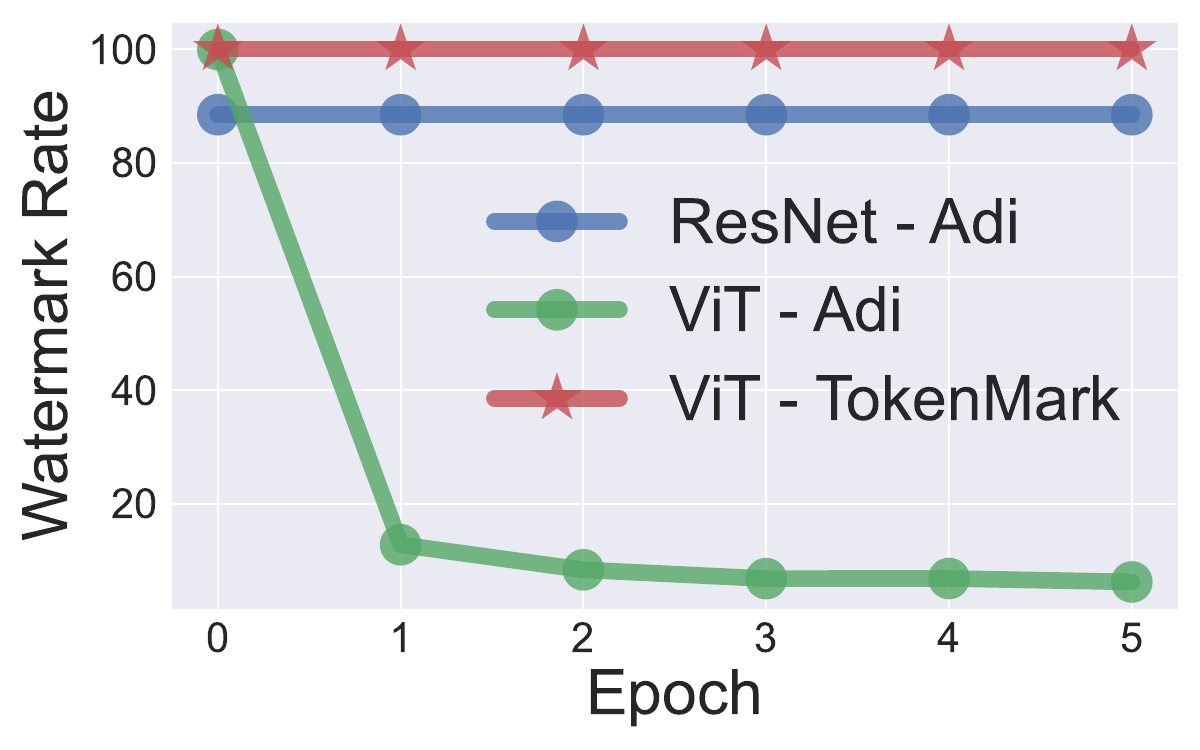}
	\caption{Robustness against fine-tuning attacks to backdoor-based watermarking on representative CNN and Transformer models, e.g., ResNet and ViT. Adi suggests the method of \cite{adi2018turning} and other setup is provided in Appendix \ref{sec:app_exp}.}
	\label{fig:cnnvsvit}
\end{figure}

Robustness poses another significant challenge to the watermarking of the pre-trained models, especially Transformer-based ones. Primarily, previous watermarking schemes mostly rely on the backdoor of the model; however, it is reported that these backdoors are not robust across all models, e.g., backdoors of the Transformers are more susceptible to removal attacks \cite{doan2023defending,yuan2023you} than backdoors of the convolutional neural networks (CNNs). We confirm this point in Fig.~\ref{fig:cnnvsvit} by comparing the watermarking rates over the fine-tuning of a ResNet and a ViT. It is clear that watermarks for ViT can hardly be extracted merely after 1 epoch. Further, the pre-train-and-fine-tune paradigm makes the pre-trained model naturally vulnerable to a variety of attacks including but not limited to fine-tuning, pruning, quantization, extraction attacks, etc. In particular, backdoor-based watermarking often faces the threat of trigger exposure in the extraction process and thus suffers from spoofing attacks. Hence the challenge is severe for pre-trained models to defend against all these attacks.

We summarize the challenges as follows: \textit{first}, many pre-trained models blur the boundary between modalities, which is in contrast to the existing modality-specific watermarking schemes. \textit{Second}, the mainstream watermarking scheme depends on backdoors which are not robust on Transformer-based models. These challenges lead to our question: \textit{can we design a modality-agnostic, task-independent, robust watermarking scheme for pre-trained models?}

As we observe, a large proportion of the state-of-the-art pre-trained models have Transformers as their backbones. It is a salient attribute of Transformer-based models to be permutation-equivariant \cite{lee2019set,naseer2021intriguing,2304.07735}, meaning that the outputs of the model can be equivalently computed on  permuted inputs. Motivated by this attribute, we consider it possible to build in watermarking as a secondary functionality independent from the main one in pre-trained models, within the same set of model weights --- the secondary function works on the specifically permuted, rather than the normal inputs. 

%However, the shuffling invariance in previous works is insufficient to support our design: it either requires modification to the original model \cite{lee2019set}, or provides no guarantee but vague robustness against shuffling \cite{naseer2021intriguing}. It is our contribution to utilizing the permutation equivariance property of Transformers \cite{2304.07735} in both the forward and backward propagation, suggesting that the model could be equivalently trained on permuted input features. This unique finding offers us a feasible way to embed two sets of parameters, i.e., the original $\theta$ and its permuted version $P(\theta)$, in one backbone model yet without interfering each other. The property allows backdoors to be embedded despite the specific modality or form of the data.

%The property warrants that the watermark can be extracted with 100\% accuracy, yet not affecting the main functionality of the model.

By this observation, we tackle the first challenge by implanting watermark as the secondary function in pre-trained models, utilizing the permutation equivariance property. Note that our scheme, \textit{TokenMark}, is significantly different from \cite{2304.07735} in that we build two sets of weights in one model but their work only uses one set of weights: the additional set of weights are obtained by fine-tuning the original model on a set of permuted samples. These samples are randomly chosen from \textit{any} dataset of \textit{any} type but permuted by a secret permutation matrix $\mP$. By the permutation equivariance property, the intermediate features are correspondingly permuted, and are restored to the correct order by $\mP^{-1}$ to produce the final outputs. Meanwhile, the main functionality of the model is preserved when being fed normal inputs. With two sets of weights, the watermarking functionality of the model is only triggered given inputs permuted by $\mP$, despite the type of the input. Hence our scheme is modality-agnostic and is independent of any downstreaming task. The comparison of the application range of different watermarking schemes is found in the leftmost of Fig.~\ref{fig:thumbnail}.

Benefited from `structural embedding,' our permutation-driven watermarking scheme is robust against various attacks. We found our `structural watermark' is fast to embed (rightmost of Fig.~\ref{fig:thumbnail}), but slow to remove indicating stronger robustness (middle of Fig.~\ref{fig:thumbnail}). It is intuitive that in backdoor-based watermarking, the backdoor samples merely concern a few neurons which trigger the targeted output; while in \textit{TokenMark}, almost all the model weights participate in permutation and watermark embedding. Hence the watermark persists through weights alterations by fine-tuning, pruning, quantization and extraction attacks. Since no trigger is involved, \textit{TokenMark} does not need to retain any trigger set and faces no threat of trigger exposure if the permutation matrix is kept secret. Meanwhile, guessing the correct permutation order is almost impossible due to the huge permutation space which is as large as the factorial number of the token dimension of a pre-trained model.

\textit{TokenMark} can flexibly serve as a plugin to various model watermarking systems. The watermarking functionality can be implemented by any means. For example, it can be a classification task as in Adi et al. \cite{adi2018turning}, or secret vector alignment in SSLGuard \cite{cong2022sslguard}. One only needs to replace the backdoor-based trigger with permuted inputs to embed the watermark, which yet takes fewer epochs than the trigger to embed. \textit{TokenMark} can be considered an enhancement to all trigger-based watermarking schemes, turning the existing schemes into more general, robust, and efficient ones.

Highlights of our contribution are as follows: we propose a new watermarking scheme \textit{TokenMark} for pre-trained models, which is independent of the data modality, the downstream task, and even the trigger set. We show that \textit{TokenMark} can serve as a universal plugin to various watermarking systems and significantly enhances their robustness. We conduct experiments on five state-of-the-art pre-trained models for CV and NLP. Experimental results have verified the superiority of \textit{TokenMark} over other schemes and revealed its potential to unify different intellectual property (IP) protection services into a single one for model ownership management.

\section{Related Works}
\label{sec:related}

Deep model IP protection has been a trendy topic in recent years. We focus on the watermarking schemes in this paper. Despite that many watermarking schemes proposed for CV models and NLP models, there are few on multi-modal ones.

\textbf{CV models.} Early works \cite{adi2018turning,zhang2018protecting} leverage model backdoor for watermarking. Trigger samples are created by overlaying trigger patterns (e.g., a white square covering a part of the image) onto normal images. The model is trained to produce a targeted output on the trigger samples. Frontier-Stitching \cite{le2020adversarial} uses adversarial examples as watermark keys. Entangled Watermarking Embedding (EWE) \cite{jia2021entangled} trains the model by a soft nearest-neighbor loss to entangle the feature representation of the watermark and training data. Free-finetuning \cite{wang2023free} incorporates a proprietary model independently trained on a custom dataset with sample-label pairs for embedding and, thus is freed from fine-tuning. Kim et al. \cite{kim2023margin} enhances robustness against extraction attacks by ensuring the predictions made by the original and the surrogate model on the trigger set are the same. The aforementioned works are designed for image classifiers and some \cite{le2020adversarial, kim2023margin} even directly alter the classifiers. But such alteration is not expected due to the fidelity requirement of watermarking.

\textbf{NLP models.}  Watermarking for large language models or their APIs has drawn much attention recently. CATER \cite{he2022cater} proposes an optimization method to decide watermarking rules that can minimize the distortion of overall word distributions while maximizing the change of conditional word selections. Kirchenbauer et al. \cite{kirchenbauer2023watermark} embed watermark by selecting a randomized set of `green' tokens before a word is generated, and then softly promoting the use of green tokens during sampling. A statistical test is adopted for detecting the watermark. In large language model-based code generation, Li et al. \cite{li2023protecting} embed watermarks by replacing tokens with synonyms. These methods typically modify the model outputs for watermark embedding.

\textbf{Pre-trained model} watermarking is less explored in the literature. SSLGuard \cite{cong2022sslguard} claims to be the first work to watermark the pre-trained backbone, but it is still optimization-based with an image trigger, and the trigger set is required to be kept secret. The method is barely applicable to other modalities.

\section{Preliminaries}
% \label{sec:background}
\label{sec:preliminary}

\subsection{Watermarking Requirements}
\label{sec:require}
Watermarking is a technique to embed a message representing the identifier into the carrier, and later to extract the message for ownership verification. With the deployment of various deep learning models these days, it has become a critical step to embed identifiable marks into a pre-trained model for protecting the intellectual property of the proprietary model. Generally, a watermarking scheme should satisfy the following properties according to \cite{lukas2022sok}:

% The terminology is from an SoK from S&P'22
$\diamondsuit$ \textbf{Effectiveness \& Integrity}. Watermarks can be extracted from the watermarked model effectively, and models trained without access to the source model do not retain the watermark.

$\diamondsuit$ \textbf{Efficiency}. The embedding and extraction process should be efficient.

$\diamondsuit$ \textbf{Fidelity}. The impact of the watermark on the model's performance should be minimal. For pre-trained models, the watermark should not affect the performance of the downstream tasks.

$\diamondsuit$ \textbf{Robustness}. The watermark should survive various model extraction and watermark removal attacks, including model extraction, fine-tuning, pruning, quantization, and even adaptive attacks where the attackers are aware of the watermarking strategy and specifically take countermoves.

%$\diamondsuit$\textbf{Undetectability}. The watermark should be undetectable by the adversary who does not have knowledge of the secret key.

Additionally, we propose a new requirement specifically for the multi-modal pre-trained model watermarking:

$\diamondsuit$ \textbf{Generality}. A watermarking method should be general enough to apply to data of a wide range of modalities. The watermarking procedure should be independent of any downstream task.

\subsection{Backdoor-based Watermarking}
\label{sec:baseline}
Backdoor-based methods have been prominent approaches for model watermarking, since they have the minimum requirement of blackbox access to the model. Most of them, such as \cite{zhang2018protecting, le2020adversarial, jia2021entangled, wang2023free, kim2023margin,cong2022sslguard}, share a form of backdoor embedding and extraction. In the embedding phase, a designed trigger, which could be randomly generated or adversarially trained, is superimposed on a small set of inputs to produce the triggered inputs. These trigger sample $x$ is paired with the targeted output label $y_t$ or distribution. To embed a watermark, the model $f$ is trained or fine-tuned both on the normal dataset and the trigger sample set. In the extraction phase, the owner feeds the trigger sample set $D$ to the suspected model to see if it could yield the targeted output labels or if the output follows the targeted distribution. The performance of watermark extraction is evaluated by the watermark rate (WR), defined by the proportion of successfully triggered inputs in the trigger sample set $D$:
\begin{equation}\label{eq:wr}
	% \mathrm{WR} = \frac{1}{|D|}\sum_{x \in D}\mathbbm{1}[\textrm{sim}(G(F(\mZ, P(\theta_*))), sk) > \epsilon_{wm}]
	\mathrm{WR} = \frac{| \{x \in D | f(x) = y_{t}\} |}{|D|}.
\end{equation}

Represented by Adi et al. in \cite{adi2018turning}, most of the backdoor-based watermarking schemes \cite{zhang2018protecting, le2020adversarial, jia2021entangled, wang2023free, kim2023margin} are designed as a classification function which predicts the targeted label given the trigger samples. It is clear that the WR of watermarking task is concerned with the proportion of trigger samples in the embedding phase, and it is important to balance the model's performance on normal samples (main task) and the triggered samples (watermarking task). 

Represented by SSLGuard \cite{cong2022sslguard}, the watermarking functionality is implemented as a vector alignment task for pre-trained models. For the triggered inputs, SSLGuard selects a random secret vector $sk$ as the target output and projects any triggered output by the decoder $G$ to the space of $sk$. The triggering process is considered successful given that the cosine similarity between the triggered output and $sk$ surpasses a given threshold $\epsilon_{wm}$. Hence the WR is defined correspondingly:
\begin{equation}\label{eq:wr_ssl}
	\mathrm{WR} = \frac{|\{x \in D | \mathrm{sim}(sk, G(f(x))) > \epsilon_{wm}\}|}{|D|},
\end{equation}
The embedding process involves training over the triggered samples with a contrastive loss which discriminates the triggered samples from the normal ones.

\subsection{Permutation Properties of Transformers} 
\label{subsec:permutation}
Permutation invariance refers to the property that the output value for a given set is the same regardless of the order of objects in the set. Deep sets \cite{zaheer2017deep} derive the necessary and sufficient conditions for permutation invariance in deep models. Set transformer \cite{lee2019set} first establishes the conditions of permutation invariance on Transformers, which has later seen applications in \cite{naseer2021intriguing, patchshuffling, engel2021point, tang2021sensory, lu2023pinat}. It is empirically verified that vision transformers are more robust to such patch shuffling, compared to convolutional networks~\cite{naseer2021intriguing}. %Other applications of this invariance property include point cloud processing \cite{engel2021point}, reinforcement learning \cite{tang2021sensory}, neural architecture search \cite{lu2023pinat}, etc. 

While previous works have primarily focused on the inter-token permutation of Transformer models, \cite{2304.07735} further demonstrates that Transformers also possess intra-token permutation equivariance. In the training process of a Transformer, if the input tokens are permuted and the outputs of the Transformer backbone is permuted back, it would be equivalent to training a model on normal inputs, but with a new set of weights correspondingly permuted. This property implies that by permuting the inputs, one can `reprogram' the model as if its weights were shuffled by the permutation. Our work utilizes the permutation equivariance of Transformers but has a different expression and application from that in \cite{2304.07735}, which we will elaborate in Sec.~\ref{sec:method}.

\section{Use Case and Threat Model}
\label{sec:threatmodel}

In this section, we first introduce the user cases of \textit{TokenMark}, followed by the threat model and the adversary's capabilities.

\textbf{Use case.} As a watermarking scheme, \textit{TokenMark} serves to provide proof of ownership for a proprietary model. Corporations often pre-train a large model on their proprietary or public data, and release the APIs to the public, or publish parameters of the model. In the API case, anyone can only query the model as a service (or EaaS, encoder as a service) \cite{eaas} whereas anyone has full access to the model parameter in the latter one. In either case, the owner of the pre-trained backbone model would often face the threats to its ownership. An adversary may steal the pre-trained model by any means and claim false ownership over the model. Worse still, the adversary might distribute a commercial version of the model for benefit.

% TokenMark现在作为一种增强方法，依附于baseline的use case，是否还要详细说明自己的验证场景？好像因为P^-1的存在，还是要依赖authority来进行水印提取
% 但是从NDSS的review来看，这个故事有些复杂，带来了很多问题
% 暂且注释掉
% To prevent dispute on its model, the owner submits a secret tuple to a trustworthy authority as a proof of its model. Such a proof cannot be easily faked unless the prover has really trained its model with the secret tuple. In resolving an ownership dispute of a target model, despite API or parameter access, the authority verifies the target model against the proof. If it could successfully extract a pre-defined high-dimensional feature vector (a part of the proof) from the model on a pre-selected permutation matrix (another part of the proof), the authority considers the model is originated from the proof provider (the owner) with high probabilities. Our watermarking system does not rely on any trigger set so that the authority can feed any valid input into the pre-trained target model for verification. Our system also guarantees that permuting the input with the permutation matrix would produce the pre-defined feature vector on the corresponding watermarked model, while outputting nonsense on unwatermarked models.

Depending on the forms of accesses, we consider three representative \textbf{threat settings} to our watermarking scheme. The goal of the adversary is to obtain a copy of the normal-functioning model while getting rid of the embedded watermark. We elaborate on the threat settings as follows.

\textbf{Black-box adversaries} merely have API accesses to the victim's model, plus some reasonable additional knowledge of the model including the model structure (Transformer) and the existence of the watermark, etc. The adversary attempts to extract the functionality of the victim's model by training a substitute on a dataset consisting of paired input queries and outputs by APIs, and it expects the substitute model does not carry any original watermark.

\textbf{White-box adversaries} have full access to the model and can modify the backbone parameters freely. We consider the adversary can perform fine-tuning, pruning, and quantization attacks, referring to the adversary further training the pre-trained model, pruning partial weights of the model, and compressing model parameters to a lower bit representation, respectively. The purpose of the adversary is to remove the original watermark by tweaking the model without affecting its main functionality.

Additionally, we consider an \textbf{adaptive attacker} who is fully aware of our watermarking scheme and designs its attack correspondingly to overwrite or to remove the watermark from the model. Some of the adaptive attackers are not realistic in the real world but act beyond the adversary to see how the watermarking system defends them.

\section{Methodology}
\label{sec:method}

% motivation
% We are motivated to design a watermarking system for pre-trained Transformers, which satisfies the requirements in Sec.~\ref{sec:require}. Our method of watermarking is inspired by the observation that the Transformer-based models are permutation equivariant \cite{2304.07735} --- the weights of the Transformer encoder can be equivalently learned on permuted inputs, which we refer to as permuted weights.  It reveals the potential of the Transformer to incorporate more than one set of parameters within one model: one set is for conducting the normal task while other sets are permuted sets which could be used for other tasks such as watermark embedding and extraction. However, the permuted set of parameters would not function unless the model being fed a correspondingly permuted input. Thus our goal is to build two sets of parameters in the pre-trained model, with one set hidden by a secret permutation. In verification, the watermark can be extracted with correctly permuted inputs to the target model.

We argue that previous backdoor-based watermarking schemes are vulnerable to various attacks mostly due to the lack of intertwinement between the main function and the watermark function within one model, which leaves room for the adversary to decouple the two, i.e., removing the watermark but keeping the main functionality of the model. Inspired by this, we aim to design a method to tightly intertwine the two functionalities within one model. 

We observe that the permutation equivariance property of Transformer-based models can be novelly applied to implementing two different functionalities within one set of weights, \textit{explicitly} --- meaning each has its own set of weights. Note that this is rare in previous backdoor-based schemes where the watermarking functionality is \textit{implicitly} built in the model weights. To explain it clearly, we begin with our perspective on permutation equivariance, and then elaborate how we utilize the property to design \textit{TokenMark}.

\subsection{Permutation Equivariance }
\label{subsec:preliminaries}
According to \cite{2304.07735}, the permutation equivariance property of Transformer-based models holds in the forward and backward propagations. Let $\mZ \in \R^{n \times d}$ be the output of the Transformer embedding layer, where $n, d$ represent the number of tokens and the dimension of each token, respectively. We also refer to $\mZ$ as the input feature to the Transformer backbone $F$ which is a stack of Transformer encoders and decoders. The weights of a Transformer encoder/decoder are $\theta = \{ \mW^{(Q)}, \mW^{(K)}, \mW^{(V)}, \mW^{(a)}, \mW^{(1)}, \mW^{(2)}, b, \gamma \}$ which are the parameters of the QKV projection matrix $Q, K, V$, attention projection, the first and second linear layer in MLP, the bias, and layer normalization parameter, respectively. Let the permutation matrix be $\mP \in \R^{d \times d}$ where each row and each column is a one-hot vector and $\mP^{-1} = \mP^{T} $. Hence the forward permutation equivariance is stated as:

\begin{theorem}[Forward equivariance]
    \label{thm:forward}
    Transformer backbone $F(\cdot)$ is permutation-equivariant in the forward propagation, i.e.,
    \begin{equation}\label{eq:forward_equivariance}
        F(\mZ\mP, \theta)\mP^{-1} = F(\mZ, P(\theta)),
    \end{equation}
where $\theta,P(\theta)$ denotes the original and permuted model weights, respectively.
\end{theorem}

Specifically, $P(\theta)$ includes the permuted version of each weight, denoted by a subscript $_P$:
\begin{align}
    \label{eq:permutation_specific}
    &\mW_P^{(i)} =\mP \mW^{(i)} \mP^{-1}, ~i = Q, K, V, a,\\
    &\mW_P^{(1)} = \mW^{(1)} \mP^{-1}, \mW_P^{(2)} = \mP \mW^{(2)},\\
    &b_P = b \mP^{-1}, \gamma_P = \gamma \mP^{-1}. \label{eq:permutation_s}
\end{align}

Note that Thm.~\ref{thm:forward} and the forward permutation equivariance property in \cite{2304.07735} are essentially different expressions of the same property, thus leading to different application scenes. In \cite{2304.07735}, the property is expressed as
\begin{equation}
	\label{eq:forward_equivariance_CVPR}
	F(\mZ\mP, P(\theta))\mP^{-1} = F(\mZ, \theta),
\end{equation}
suggesting the model weights can be equivalently used for inference under permutation, i.e., $P(\theta)$ is used instead of $\theta$ in $F(\cdot)$. This form supports applications including model encryption, privacy-preserving split learning \cite{2304.07735}, etc. In contrast, our expression of Eq.~(\ref{eq:forward_equivariance}) indicates that having the permuted inputs forwarded on $\theta$ is equivalent to having the original inputs forwarded on $P(\theta)$. This is useful as $P(\theta)$ could act as the second set of weights under the hood of $\theta$, thereby serving as the weights for watermarking.

What makes it even better is that the backward permutation equivariance holds as well:
\begin{theorem}[Backward equivariance \cite{2304.07735}]
    \label{thm:backward}
    Transformer backbone $F(\cdot)$ is permutation-equivariant in the backward propagation w.r.t. loss $\ell$, i.e.,
        \begin{align}
            &\frac{\partial \ell}{\partial \mW_P^{(i)}} =\mP \frac{\partial \ell}{\partial \mW^{(i)}} \mP^{-1}, ~i = Q, K, V, a,\\
            &\frac{\partial \ell }{\partial \mW_P^{(1)}} = \frac{\partial \ell}{\partial \mW^{(1)}} \mP^{-1}, \frac{\partial \ell}{\partial \mW_P^{(2)}} = \mP \frac{\partial \ell}{\partial \mW^{(2)}},\\
            &\frac{\partial \ell }{\partial b_P} = \frac{\partial \ell }{\partial b} \mP^{-1}, \frac{\partial \ell }{\partial \gamma_P} = \frac{\partial \ell }{\partial \gamma} \mP^{-1}.
        \end{align}
\end{theorem}

Combining Thm.~\ref{thm:forward} and Thm.~\ref{thm:backward}, one could figure out that the Transformer backbone is permutation-equivariant throughout the training by recursion. At the end of a single iteration (one forwarding followed by a backward propagation), the resulted model weights $\theta_*$ could relate to the specifically permuted weights $P(\theta_*)$ as follows:
\begin{equation}
    \label{eq:train_equivariance}
        F(\mZ\mP, \theta_*)\mP^{-1} = F(\mZ, P(\theta_*)),
\end{equation}
meaning that if we feed permuted input feature $\mZ \mP$ into $F$ and reverse the permutation on the output feature of $F$, the set of parameters $\theta_*$ would function as $P(\theta_*)$ instead of $\theta_*$. Hence by the forward and backward equivariance property, we know that the weights of $F(\cdot)$ trained on permuted inputs is equivalent to the permuted weights trained on normal inputs. The latter can be the second set of weights apart from $\theta$.

\subsection{Design of TokenMark}
\label{subsec:overview}
\textbf{Motivation.} Eq.~(\ref{eq:train_equivariance}) reveals that training on permuted inputs produces a set of permuted weights; meanwhile, Eq.~(\ref{eq:forward_equivariance}) suggests that inference on permuted inputs produces the same results of inference on normal inputs and permuted weights. The former can be applied in watermark embedding with the permuted inputs serving as backdoors. The latter can be adopted in watermark extraction: the correspondingly permuted $\mZ\mP$ can produce the targeted output on $\theta_*$ but not on other models; neither the normal ones $\mZ$ nor the wrongly permuted ones $\mZ\mP'$ can be triggered on $\theta_*$.

Our motivation is beyond a new form of backdoor triggers in watermarking. From the weights' perspective, the embedding of the conventional trigger-based samples hinges upon the excessive capacity of $\theta_*$ to incorporate the watermark function beyond the main one, while the permuted backdoor samples build in an explicit set of weights $P(\theta_*)$ which keeps the original $\theta_*$ almost intact, thereby not impacting the main function. Hence the permuted backdoor samples\textit{ take less efforts to embed} than conventional ones. What is more important, our backdoor watermark takes the entirety of the model weights to embed, which is more tightly associated with the model weights than the traditional backdoor trigger which is merely concerned with some `neuron path' of the model. Therefore, the permuted watermark is \textit{more resilient }against various attacks including fine-tuning, pruning, quantization, extraction, etc. The attacker has to destroy a large proportion of $\theta_*$ to remove the watermark, which however hurts the main functionality of the model leading to a failed attack. Finally, permutation has nothing to do with the specific modality of the input, which makes the watermarking scheme \textit{universal} across different models.

In summary, the incentive behind \textit{TokenMark} is an innovative exploit of the permutation equivariance property: the permuted backdoor is embedded with an explicit set of weights, which is more robust, and universal across different modalities. 

\textbf{Goal.} For its design, we set the fidelity and effective goals for \textit{TokenMark}:
\begin{definition}[Fidelity]
    \label{def:model_ultility}
    For any downstream head $DS(\cdot)$, the loss of the watermarked weights $\theta_*$ should be close to that of the original ones $\theta$:
    \begin{equation}\label{eq:fidelity}
        \Theta = \{\theta_* | ~~\|  \Ls_{DS}(F(\mZ, \theta_*)) - \Ls_{DS}(F(\mZ, \theta)) \|_2 < \epsilon_{DS}\},
    \end{equation}
    where $\Theta$ is the set of pre-trained backbone parameters that are useful.
\end{definition}

From the watermarking aspect, we expect the permuted weights to accomplish the watermarking task by outputting targeted labels when triggered:
\begin{definition}[Effectiveness]
    \label{def:watermarked_parameters}
    The watermarked weights $\theta_*$ should yield the targeted output by decoder $G(\cdot)$ with a small loss:
    \begin{align} \label{eq:effect}
        \Theta_P = \{\theta_*\in\Theta | &\Ls_{wm}(G(F(\mZ,\mathrm{P}(\theta_*))), y_t) < \epsilon_{wm}\\\wedge &\Ls_{wm}(G(F(\mZ,\theta_*)), y_t) > \epsilon_{wm}\},
    \end{align}
    where $\Theta_P$ is the set of watermarked parameters, and $y_t$ is the targeted output which varies according to the watermark function.
\end{definition}

\begin{figure}
    \centering
    \includegraphics[width=1.0\linewidth]{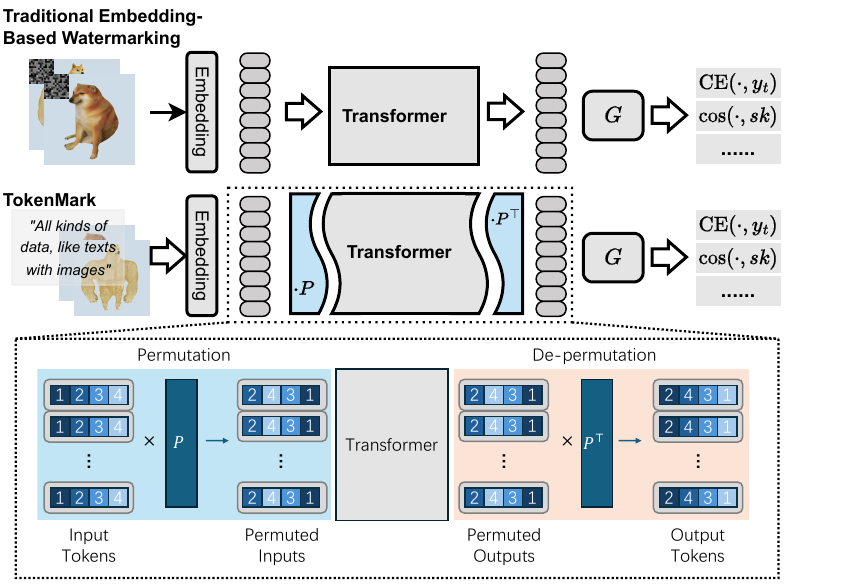}
    \caption{Comparison between the traditional watermarking scheme and TokenMark.}
\label{fig:overview}
\end{figure}

\textbf{Overview.} \textit{TokenMark} consists of two phases --- the embedding and the extraction. Prior to embedding, the owner obtains the original model $\theta$ following the normal training procedure. 

At the embedding stage, $\theta$ is fine-tuned on a very small `embedding set' that contains arbitrarily chosen samples from any dataset. To complete the watermark task, each sample from the embedding set is permuted by a secret permutation matrix $\mP$ before being fed into the model, with its output feature being first restored by $\mP^{-1}$ and then fed into the decoder $G$ to generate the final output, as depicted by Fig.~\ref{fig:overview}. According to Eq.~(\ref{eq:train_equivariance}), the `permutation-de-permutation' paradigm results in two sets of parameters in one model: $\theta_*$ and $P(\theta_*)$. While $\theta_*$ functions as original $\theta$, $P(\theta_*)$ can complete the watermark task. Done embedding, the decoder $G$ along with the permutation matrix $\mP$ are kept secret. In extraction, the permuted extraction set is fed to the model for verification. Note that both the embedding and extraction sets are different from the trigger set in the backdoor-based watermarking: there is no strict requirement on the quantity, content, or form of the data, nor any modification to the input. The extraction set could be totally different from the embedding set. 

In the normal mode, the pre-trained model $\theta_*$ works just as any other pre-trained model. When a suspected model is observed, by feeding the model with permuted inputs and de-permuting the outputs, the restored output feature can be decoded by the decoder $G$ to extract the pre-defined watermark. The watermarking task can be any in the previous backdoor-based watermarking methods. For example, the classification one in Adi et al. \cite{adi2018turning}, or the vector alignment in SSLGuard \cite{cong2022sslguard}. In the following section, we choose two representative watermarking methods in Sec.~\ref{sec:baseline} to illustrate how \textit{TokenMark} hardens them.

\subsection{Backdoor Watermarking: TokenMark-B}
\label{subsec:TokenMark-b}
To show how TokenMark enhances conventional backdoor-based watermarking for models, we build TokenMark-B upon Adi et al.~\cite{adi2018turning} where the watermark task is to classify backdoored samples to the target label.

The \textbf{embedding} of TokenMark-B is similar to the embedding procedure described in Sec.~\ref{sec:baseline} except that the input feature of the backdoor sample ($\mZ$) is permuted instead of the input ($x$) being directly patched. The part of permuted inputs are trained with the cross entropy loss as follows:
\begin{equation}
    \label{eq:TokenMark-B_effectiveness}
    \Ls_{wm} = \mathrm{CE}(G(F(\mZ\mP, \theta_*)\mP^{-1}), y_t)
\end{equation}
where $y_t$ is the target label. Any downstreaming task is trained normally: 
\begin{equation}
    \label{eq:TokenMark-B_fidelity}
    \Ls_{DS} = \mathrm{CE}(DS(F(\mZ, \theta_*)), y)
\end{equation}
where $y$ is the ground-truth label. The decoder $G$ is set to be a random mapping which is kept secret for the owner.

From our practical experience, few- or even one-shot learning is typically sufficient for embedding TokenMark-B. Hence we only update the pre-trained model with Eq.~(\ref{eq:TokenMark-B_effectiveness}) for a few epochs to embed the watermark.

The \textbf{extraction} of TokenMark-B is to decode the target label with samples permuted by $\mP$ from the watermarked model. The metric WR defined in Eq.~(\ref{eq:wr}) is used to evaluate extraction performance.

\subsection{ SSL Watermarking: TokenMark-S}
\label{subsec:TokenMark-s}

To demonstrate how TokenMark improves the watermarking in self-supervised learning (SSL) setting, we propose TokenMark-S based on SSLGuard \cite{cong2022sslguard} where the watermark task is to extract a pre-defined secret vector $sk$ from the model.

The \textbf{embedding} of TokenMark-S considers two aspects: one for the fidelity of the pre-trained model and the other for the effectiveness in extracting watermarks. For fidelity, we aim to keep the functionality of the original pre-trained model, yet without prior knowledge of any downstream task. Similar to SSLGuard, we try to match the output of $\theta_*$ with that of $\theta$ by minimizing the loss:
\begin{equation}\label{eq:loss_match}
    \Ls_{DS}(\theta, \theta_*) = -\mathrm{sim}(F(\mZ, \theta), F(\mZ, \theta_*)),
\end{equation}
where $\mathrm{sim}(\cdot, \cdot)$ denotes the cosine similarity between two output feature vectors.

For effectiveness, we define a pair of losses as $\Ls_{wm}$ which includes a correlation loss on the permuted inputs as well as an uncorrelated loss on the normal inputs:
\begin{align}
\Ls_{corr}(sk, G(F(\mZ, P(\theta_*)))) &= - \mathrm{sim}(sk, G(F(\mZ, P(\theta_*)))),\\ 
\Ls_{uncorr}(sk, G(F(\mZ, \theta_*))) &= \mathrm{sim}(sk, G(F(\mZ, \theta_*)))^2.
\end{align}
Minimizing $\Ls_{uncorr}$ would push the cosine similarity to zero and thus the two vectors are close to orthogonal. 

Hence the overall embedding loss for the pre-trained model is:
\begin{align}\label{eq:loss_backbone}
    \Ls_{B} 
    = \Ls_{DS}(\theta, \theta_*) 
    +& \Ls_{corr}(sk, G(F(\mZ, P(\theta_*)))) \\ \nonumber
    + &\Ls_{uncorr}(sk, G(F(\mZ, \theta_*))).
\end{align}

The shadow model of SSLGuard for robustness is kept in TokenMark-S. The shadow model $\theta_s$ of the same structure with $\theta$ is trained by $\Ls_{S}$ following the same form to $\Ls_{DS}$ in Eq.~(\ref{eq:loss_match}). Meanwhile, the decoder $G$ is trained to extract the secret vector $sk$ from the watermarked model $P(\theta_*)$. Taking the shadow model into account, $sk$ should be extracted from $\theta_s$ as well. Moreover, the secret vector should not be decoded from the unwatermarked model $\theta$, or the watermarked model with incorrect permutation matrix $\tilde{\mP}$. Hence the loss function for the decoder is defined as:
\begin{align}\label{eq:loss_decoder}
    \Ls_{G} 
    =& \Ls_{corr}(sk, G(F(\mZ, P(\theta_*)))) + \Ls_{corr}(sk, G(F(\mZ, P(\theta_s)))) \\
    +& \Ls_{uncorr}(sk, G(F(\mZ, \tilde{P}(\theta_*)))) + \Ls_{uncorr}(sk, G(F(\mZ, \theta))),
\end{align}
where $F(\mZ, \tilde{P}(\theta_*)) = F(\mZ\tilde{\mP}, \theta_*)\tilde{\mP}^{-1}$ according to the permutation equivariance property. We train $\Ls_{B}, \Ls_{G}, \Ls_{S}$ iteratively in the embedding and provide the detailed training procedures in the Appendix~\ref{sec:app_alg}.

The \textbf{extraction} of TokenMark-S is to feed the pre-trained model with an extraction set permuted by $\mP$, restore the output feature by $\mP^{-1}$, and feed the restored feature to $G$ to see how the decoded vector aligns with $sk$. Following SSLGuard, we gauge the metric WR as in Eq.~(\ref{eq:wr_ssl}) with $\epsilon_{wm} = 0.5$. By our practice, we found that the performance of TokenMark-S are consistently well even when $\epsilon_{wm}$ varies in a wide range.

\section{Discussion}
\label{sec:discussion-on-permutation}
While permutation in \textit{TokenMark} can be considered a new type of backdoor trigger, it is essentially different from the conventional trigger.

First of all, the traditional trigger-based watermarking schemes are data-driven, relying heavily on the trigger pattern and embedding set distribution. This brings two major defects for previous watermarking schemes: 1) The watermarking schemes weaken on Transformer-based models as these models are more vulnerable to backdoors \cite{doan2023defending,yuan2023you} than CNNs, leading to easy removal of the backdoor. 2) According to \cite{tan2023deep}, the extracted model may not maintain the watermark particularly when the extraction data has a different distribution from the embedding data. In contrast, TokenMark is structure-driven, thereby eliminating all the drawbacks of the conventional trigger. Since the permutation is associated with the model itself, rather than any dataset or any downstream task, our watermarking scheme is universal across different modalities.

%The permutation-inversion adversarial watermark embedding draws a clear line between $\theta_*$ and $P(\theta_*)$. The scheme combines two models in one set of parameters, one for encoding, and another for watermarking, and minimizes the interference of the two models. The permutation equivariance explains how the two models coexist and how they are optimized. In Sec.~\ref{sec:robustness}, we show that TokenMark is much more robust than the trigger-based watermarking with the same embedding method. Moreover, the structure-driven watermarking abandoned the concept of trigger set, making it more robust against the model extraction attack.

Second, it may be considered viable to reconstruct permutation by reverse engineering, but the hardness of permutation recovery is at least comparable to trigger reconstruction, if not harder. Let's assume the hidden dimension of the model is at the same order with the number of pixels of an image trigger, e.g., $d=768$ for a base-size ViT. The number of possible trigger patterns is $256^{d}$ which is orders of magnitude less than the number of possible permutations $d!$. Hence the random search attack to reverse the permutation is at least as hard as reconstructing the trigger pattern. In fact, the permutation reversing is symmetric to the trigger reconstruction as the attacker works on the model weights and inputs, respectively. The former problem is more difficult to tackle as the permutation space is discrete, while the trigger space can be continuous over which is easier to optimize by gradient-based methods.

For a deeper analysis on the permutation space, we give the following example. Taking the base-size Transformer as an example, its token dimension $d=768$ and thus the permutation matrix is of size $\R^{768\times768}$. Constrained by multi-head operation, the permutation is limited inside each head, but not across heads, which leads to a total of $12! \times 64! \approx 6.1\times10^{97}$ possible permutations (as we only consider column shuffling). It is almost impossible for an adversary to launch a brute-force attack to derive the permutation order. Further, any proprietary or commercial pre-trained model could easily have much higher dimensions than the base size. The larger the model, the more secure the watermark.

\section{Experiments}
\label{sec:experiments}
In this section, we present the experimental evidence that TokenMark is superior to the trigger-based watermarking scheme in meeting the watermarking requirements including effectiveness, integrity, efficiency, fidelity, and robustness. 

% \subsection{Experimental Setup}
% Hengyuan：由于后文Embedding Setup需要二级黑体标题，此处暂且细分出两个Setup
\subsection{Setup}
\label{subsec:setup}

\textbf{Pre-trained models}. We conduct experiments on five real-world pre-trained Transformer backbones, which are described in detail below. All models contain the base-size Transformer backbone, with token dimension $d = 768$, 12 layers, and 12 heads. Shadow model $\theta_s$ shares the same structure with the Transformer backbone.

\textit{BERT \cite{devlin2018bert}, GPT-2 \cite{radford2019gpt2}} and \textit{LLaMA 2 7B} are natural language models. BERT is a bidirectional Transformer pre-trained using a combination of masked language modeling and next sentence prediction objectives on a large corpus comprising the Toronto Book Corpus and Wikipedia. GPT-2 is a generative model pre-trained on the 40GB WebText dataset. We adopt the implementation from Huggingface Transformers for these two models. LLaMA 2 7B pre-trained by Meta serves as a representative of large language models.
%More precisely, inputs are sequences of continuous text of a certain length and the targets are the same sequence, shifting one token (word or piece of word) to the right. The model uses internally a mask mechanism to make sure the predictions for the token $i$ only use the inputs from $1$ to $i$ but not the future tokens. 

\textit{ViT-timm, ViT-Dino v2 and ViT-CLIP} are computer vision models. ViT-timm is a Vision Transformer (ViT) model trained on the ImageNet-21k image classification dataset. We select it as a representative of \emph{supervised} pre-trained vision transformers. We use the implementation from PyTorch Image Model\footnote{https://github.com/rwightman/pytorch-image-models} (timm).
Dino v2 \cite{oquab2023dinov2} is a ViT model trained with multiple self-supervised learning objectives. It is an outstanding general-purpose image embedding model and we use it as a representative of \emph{self-supervised} pre-trained vision transformers.
CLIP (Contrastive Image-Language Pretraining)~\cite{radford2021clip} is a multi-modal transformer trained on a large corpus of (image, text) pairs by aligning the similarity between the corresponding image and text embeddings. We adopt the vision encoder of CLIP as a representative of \emph{multi-modal} pre-trained vision transformers.

\begin{table}[t]
    \caption{Datasets, downstream tasks, fidelity metrics, and fine-tuning epochs.}
    \label{tab:tasks}
    \centering
    \scalebox{0.8}{\begin{tabular}{cccc}
    \hline
       Datasets     & Tasks                                                               & Fidelity Metrics                                                             &  Epochs                                                          \\ \hline
    Cifar10    & 10-classification                                                   & Accuracy                                                           &  5      \\
    STL10      & 10-classification                                                   & Accuracy                                                           &  5       \\
    CelebA     & \begin{tabular}[c]{@{}c@{}}Attributes\\ classification\end{tabular} & Accuracy                                                           &  5     \\ 
    Tiny ImageNet & 200-classification                                                & Accuracy                                                           &  5     \\\hline
    IMDB       & \begin{tabular}[c]{@{}c@{}}Binary\\ classification\end{tabular}     & Accuracy                                                           & 2                                                                \\
    SQuAD      & \begin{tabular}[c]{@{}c@{}}Question\\ answering\end{tabular}        & \begin{tabular}[c]{@{}c@{}}Exact Match (EM)\\ F1 score\end{tabular} &  3                                   \\
    SWAG       & Multiple choice                                                     & Accuracy                                                           &  3                                      \\
    WikiText-2 & Text generation                                                     & Perplexity                                                         & 1              \\ \hline
    \end{tabular}}
\end{table}

\begin{table*}[t]
	\caption{Effectiveness: WR(\%) of different pre-trained NLP models. The notations are the same with Table~\ref{tab:effectiveness}.}
	\label{tab:effectiveness_NLP}
	\centering
	\scalebox{0.8}{\begin{tabular}{ccccccc}
			\hline
			Extraction Set & \multicolumn{2}{l}{QNLI} & \multicolumn{2}{l}{IMDB}                              & \multicolumn{2}{l}{SQuAD}                                             \\ 
			Model        & BERT    & GPT-2    & BERT    & GPT-2 & BERT    & GPT-2\\ \hline
			TokenMark-B     & 100.0 (0.00)   & 100.0 (0.00)   & 100.0 (0.00)   & 100.0 (0.04)   & 100.0 (0.00)  & 100.0 (0.07)    \\
			TokenMark-S   & 100.0 (0.00)    & 100.0 (0.00)   & 100.0 (0.00)   & 100.0 (0.00)   & 100.0 (0.00)  & 100.0 (0.00)   \\
			\hline 
	\end{tabular}}
\end{table*}

\begin{table*}[t]
	\caption{Effectiveness: WR(\%) and False Positive Rate (\% in bracket) of different pre-trained CV models. False positive rates are measured on non-watermarked models.}
	\label{tab:effectiveness}
	\centering
	\scalebox{0.8}{\begin{tabular}{cccccccccc}
			\hline
			Extraction Set & \multicolumn{3}{l}{Cifar10} & \multicolumn{3}{l}{CelebA}                              & \multicolumn{3}{l}{ImageNet}                                             \\ 
			Model        & timm    & Dino V2    & CLIP    & timm    & Dino V2    & CLIP & timm    & Dino V2    & CLIP\\ \hline
			Adi et al. & 99.94 (0.26)  & 99.93 (4.91) & 99.89 (0.89)   & 89.468 (2.49) & 95.19 (2.13) & 98.80 (1.85)  & 98.68 (1.11) & 99.16 (5.06) &99.13 (1.98)\\
			TokenMark-B     & 100.0 (0.00) & 100.0 (0.00) & 100.0 (0.00) & 100.0 (0.00) & 100.0 (0.03) & 100.0 0.00() & 100.0 (0.00) & 100.0 (0.00) & 100.0 (0.00)  \\ 
			SSLGuard   & 100.0 (0.00)    & 100.0 (0.00) & 100.0 (0.00)   & 100.0 (0.00)   & 99.79 (11.2) & 100.0 (0.21)  & 100.0 (0.00)  & 99.68  (0.16) & 100.0(0.00) \\ 
			TokenMark-S   & 100.0 (0.00) & 100.0 (0.08) & 100.0 (0.00) & 100.0 (0.04) & 100.0 (0.01) & 100.0 (0.00) & 100.0 (0.00) & 100.0 (0.04) & 100.0 (0.00)\\
			\hline
	\end{tabular}}
\end{table*}

\begin{figure*}[t]
    \centering
    \begin{subfigure}[b]{0.195\linewidth}
        \includegraphics[width=\linewidth]{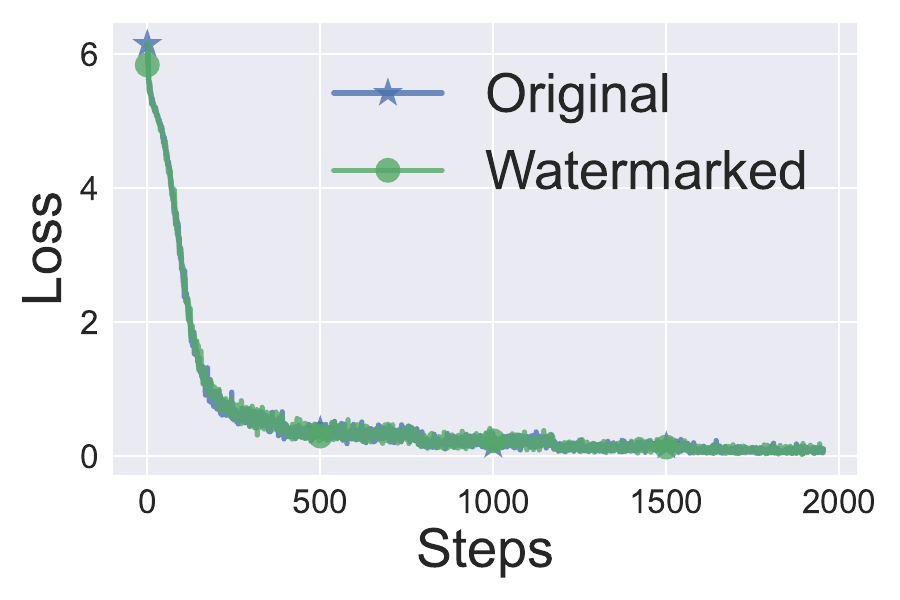}
        \caption{ViT-timm }
        \label{fig:loss_ViT_timm}
    \end{subfigure}
    \begin{subfigure}[b]{0.195\linewidth}
        \includegraphics[width=\linewidth]{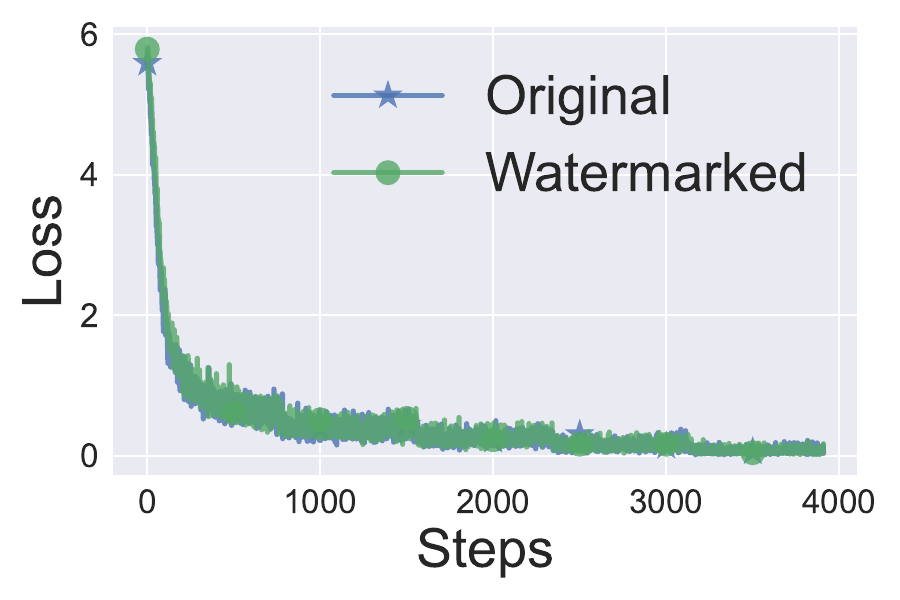}
        \caption{ViT-Dino v2 }
        \label{fig:loss_Dinov2}
    \end{subfigure}
    \begin{subfigure}[b]{0.195\linewidth}
        \includegraphics[width=\linewidth]{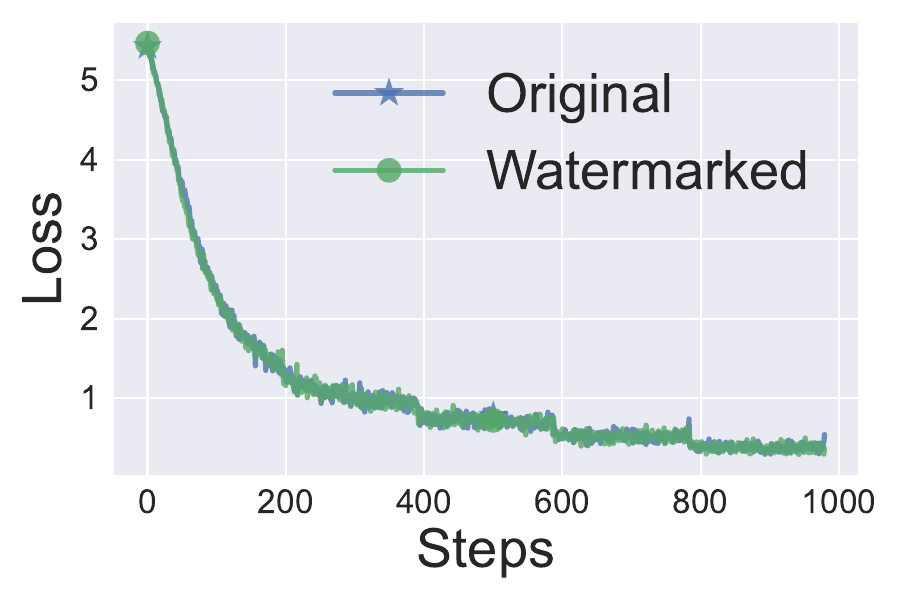}
        \caption{ViT-CLIP}
        \label{fig:loss_CLIP}
    \end{subfigure}
    \begin{subfigure}[b]{0.195\linewidth}
        \includegraphics[width=\linewidth]{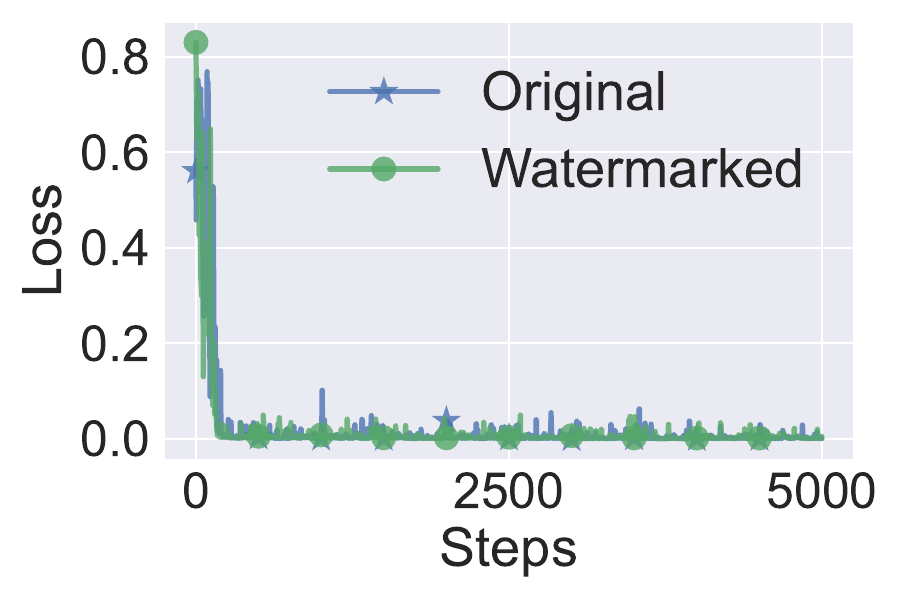}
        \caption{BERT}
        \label{fig:loss_BERT}
    \end{subfigure}
    \begin{subfigure}[b]{0.195\linewidth}
        \includegraphics[width=\linewidth]{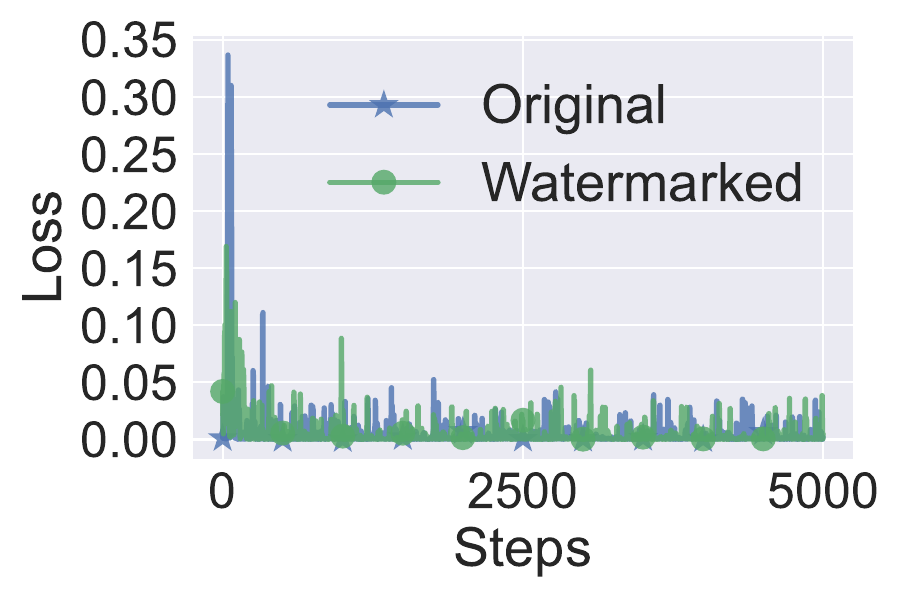}
        \caption{GPT-2}
        \label{fig:loss_GPT2}
    \end{subfigure}
    \caption{Loss curves of fine-tuning with original backbone and watermarked backbone. CV models are fine-tuned on Tiny ImageNet and NLP models are fine-tuned on IMDB.}
    \label{fig:fidelity}
\end{figure*}

\begin{figure*}[t] % have to put it here to make show up in the right place
    \centering
    \includegraphics[width=1.\linewidth]{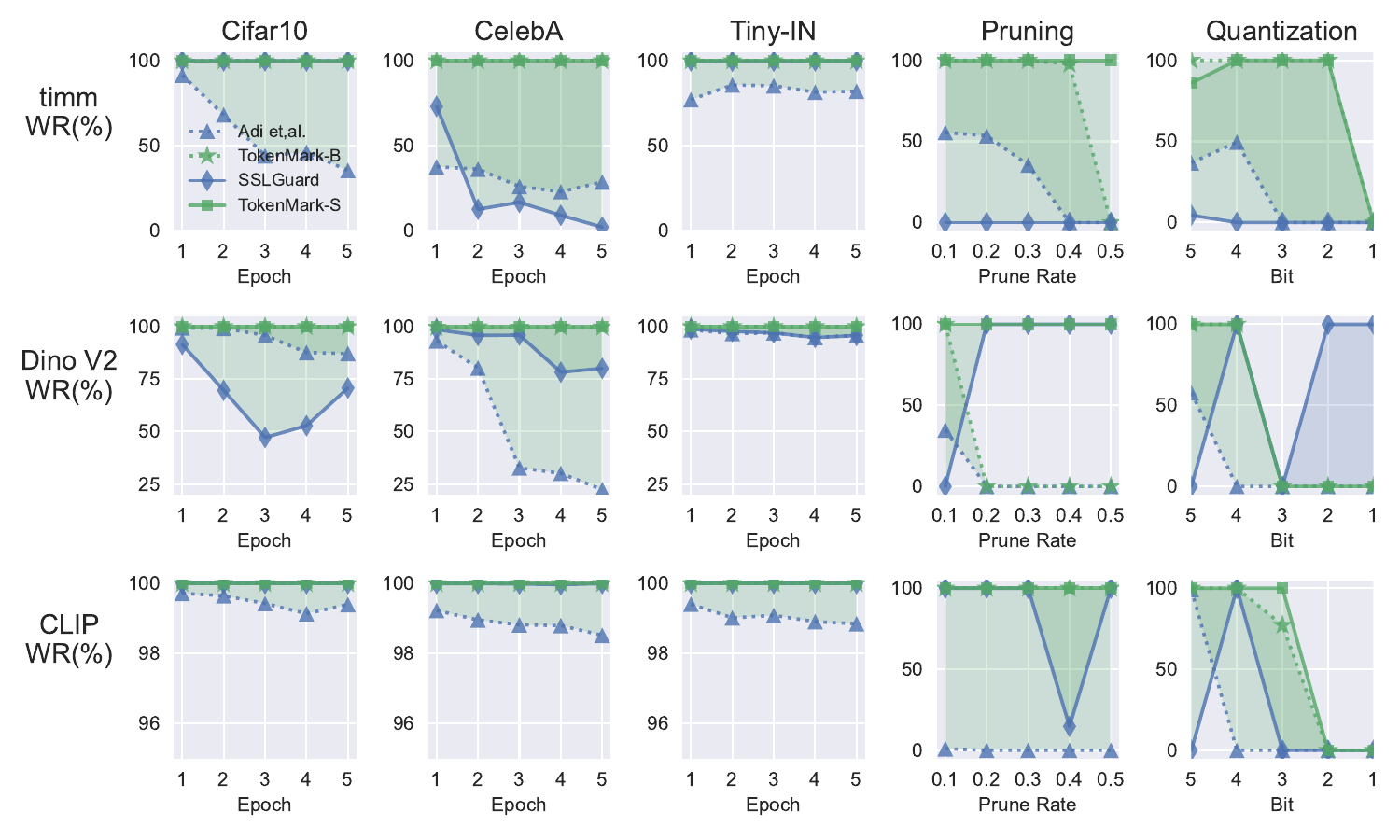}
    \caption{Performance of different watermarking schemes against white-box watermark removal attacks. Regions where TokenMark exceeds the baseline are colored green, otherwise blue. First three columns are fine-tuning attacks using different datasets, and the last two are pruning and quantization attack respectively.}
    \label{fig:white-box}
\end{figure*}

% \textbf{Datasets and tasks}. The following datasets are used in our experiments, covering a wide variety of downstream tasks from both computer vision and natural language processing.  Description and fine-tuning configurations of each dataset are provided and summarized in Table~\ref{tab:tasks}.
% Hengyuan：该部分包含数据集介绍、作用、微调超参，过长，移动到附录，主要数据集已经列在上表中了

\textbf{Datasets and tasks}. The following datasets are used in our experiments, covering a wide variety of downstream tasks from both CV and NLP. We used Cifar10 to embed watermarks into vision models and QNLI to embed NLP models. The extraction sets are test sets from Cifar10, CelebA, ImageNet, QNLI, IMDB, SQuAD, SWAG, and WikiText2. These datasets are also used for linear probing the fidelity of the watermarked models, and for evaluating robustness against the fine-tuning attack. We list the configurations in Table~\ref{tab:tasks} and more detail can be found in Appendix~\ref{sec:app_exp}.

\textbf{Embedding configurations.} We show the embedding detail of our watermarking schemes --- TokenMark-B and TokenMark-S, and baselines --- Adi et al. \cite{adi2018turning} and SSLGuard \cite{cong2022sslguard} as follows.

\textit{Embedding ViTs with Adi et al. \cite{adi2018turning} and TokenMark-B}. We select Cifar10 as the embedding set and designate 0 as the target class for triggered samples. A half of the embedding set is triggered for \cite{adi2018turning} while a set of randomly selected 256 samples is permuted for TokenMark-B. The task head is a randomly initialized linear layer from 768 to 10. The Adam optimizer with a learning rate of $10^{-5}$ is employed, and the batch size is set to 256. The scheme of Adi et al. is trained for one epoch and TokenMark-B is trained for one step. Extraction sets are test sets of Cifar10, CelebA, and ImageNet.

\textit{Embedding ViTs with SSLGuard and TokenMark-S}. Cifar10 is used as the embedding set. The Adam optimizer with learning rate $10^{-5}$ is adopted with batch size set to 16. SSLGuard is trained for 10 epochs and TokenMark-S is trained for 500 steps. Extraction sets are test sets of Cifar10, CelebA, and ImageNet. 

\textit{Embedding BERT and GPT with TokenMark}. We use 32 records from IMDB for TokenMark-B and 2,400 records from QNLI for TokenMark-S as the default embedding set. The Adam optimizer with learning rate $10^{-5}$ is employed. TokenMark-B is trained for a single training step and TokenMark-S is trained for 500 steps with batch size 8. Extraction sets are test sets from QNLI, IMDB, and SQuAD.

\textit{Embedding LLaMA 2 7B with TokenMark-B}. We use 5000 records from IMDB as the embedding set and set 10\% of them as permuted samples. An AdamW optimizer with a learning rate of $2 \times 10^{-5}$ is used. The embedding process is a few-shot full fine-tuning.

\textit{Extraction}. The watermark threshold $\epsilon_{wm}$ is set to 0.5 by default in measuring the watermark rate (WR).

\subsection{Effectiveness and Efficiency}
\label{subsec:effectiveness}
In this section, we verify the effectiveness, integrity, and efficiency performance of TokenMark. 

For \textbf{effectiveness}, we measure the watermarking rates of the models watermarked by different methods, as well as the false positive rates of different methods in detecting watermarks on the unwatermarked models . We extract watermarks using three extraction sets. Note that the original implementation of SSLGuard uses the same dataset for both embedding and extraction, but TokenMark does not have such a requirement.

\textbf{Results.} As shown in Table~\ref{tab:effectiveness} and Table~\ref{tab:effectiveness_NLP}, watermarks can be extracted with 100\% accuracy from the watermarked models by TokenMark, while shy of 100\% extraction accuracy by other methods. The high WR may be attributed to the vulnerability of Transformer models to backdoors, despite their forms. Notably, TokenMark can extract watermarks from out-of-distribution data with 100\% accuracy, e.g., in cases where the extraction set is not Cifar10. The WR shy of 100\% in these cases for Adi et al. and SSLGuard indicates these watermarking methods have some transferability to unseen extraction sets, but cannot guarantee 100\% extraction. Meanwhile, the false positive rates are almost all zeros for TokenMark but not for Adi et al. and SSLGuard, suggesting the baseline methods may wrongly detect watermarks from unwatermarked models.

To evaluate \textbf{efficiency}, we record the runtime of our experiments conducted on a single NVIDIA GeForce RTX 4090 GPU with 24GB memory. As shown on the rightmost side of Fig.~\ref{fig:thumbnail}, while Adi et al. requires 120 seconds (a training epoch) for embedding, TokenMark-B only takes 4 seconds for an effective one-shot embedding. SSLGuard takes even more time, i.e., 5.5 hours (10 training epochs) for embedding, but TokenMark-S only takes 5 minutes (500 steps).

The results show that TokenMark satisfies the basic effectiveness requirement of watermarking, and is lightweight to run.

\subsection{Fidelity}
\label{subsec:fidelity}

By Eq.~(\ref{eq:fidelity}), the fidelity of watermarking is evaluated by the gap in the fine-tuning performance between the watermarked model and the original one under the same hyper-parameter setting. The fine-tuning performance is measured by metrics listed in Table~\ref{tab:tasks}. The three vision models are fine-tuned on Cifar10, STL10, CelebA and Tiny ImageNet, and the two NLP models are fine-tuned on IMDB, SQuAD, SWAG (for BERT), and WikiText2 (for GPT), respectively. 

\textbf{Results} show that almost all the watermarked models achieve an accuracy (or, exact match and other metrics) within a variance of $\pm0.5\%$ to the clean accuracy. It confirms that all watermarking methods could well preserve the fidelity in downstream tasks. The fine-tuning loss curves of the original model and the model watermarked by TokenMark-S are presented in Fig.~\ref{fig:fidelity}. The results demonstrate that the watermark almost has no impact on the fine-tuning process. Due to space limitation, we put other results in Appendix~\ref{sec:app_fidelity}.

\subsection{Robustness}
\label{sec:robustness}
We instantiate the threat models defined in Sec.~\ref{sec:threatmodel} to evaluate the robustness of TokenMark against progressively stronger adversaries. For white-box adversaries with access to parameters of the watermarked model, we consider watermark removal via fine-tuning, pruning, and quantization. For black-box adversaries, we consider model extraction attacks under Encoder as a Service (EaaS) settings and assume they have unlimited API access to the watermarked model.  Finally, we consider an adaptive adversary with knowledge of our TokenMark scheme. We assume the goal of the adversaries is to remove the original watermark from the watermarked model or to overwrite the existing watermark.

\begin{table}[h]
	\caption{Performance under fine-tuning attack: watermark rates (\%) on fine-tuned BERT and GPT-2.}
	\label{tab:finetune-NLP}
	\centering
	\scalebox{0.8}{\begin{tabular}{ccccccc}
			\hline
			Downstream & \multicolumn{2}{l}{IMDB} & \multicolumn{2}{l}{SQuAD} & SWAG & WikiText2                                   \\ 
			Model        & BERT    & GPT-2    & BERT    & GPT-2 & BERT    & GPT-2\\ \hline
			TokenMark-B    & 100.0   & 100.0  & 100.0  & 100.0  & 100.0  & 100.0 \\
			TokenMark-S   & 100.0   & 100.0  & 100.0  & 100.0  & 100.0  & 100.0   \\
			\hline 
	\end{tabular}}
\end{table}

% fig and tab in one row
\begin{figure}[h]
	\centering
	\begin{minipage}[h]{0.45\linewidth}
		\includegraphics[width=\linewidth]{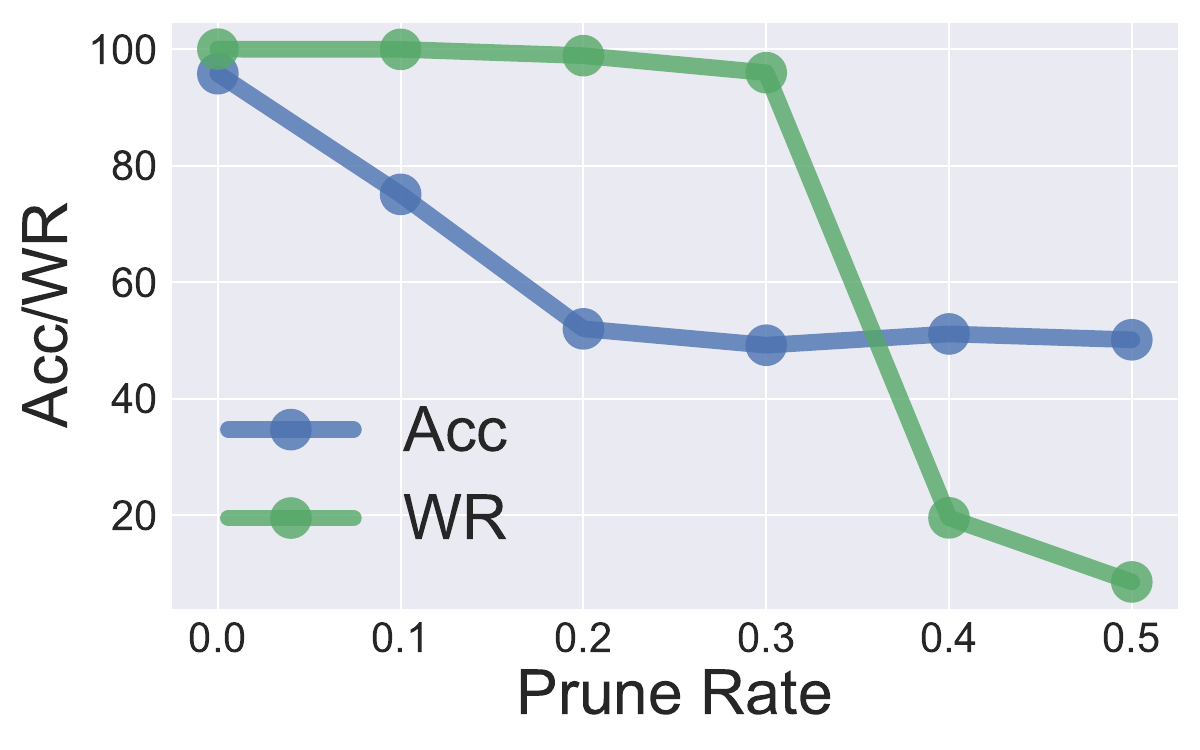}
		\caption{Performance of TokenMark-B against pruning attack on LLaMA. Acc (\%) is the accuracy of the downstream task.}
		\label{fig:llama_prune}
	\end{minipage}
	\begin{minipage}[h]{0.45\linewidth}
		\centering
		% \subcaption{Quantization}
		\captionof{table}{Performance of TokenMark-B against quantization attack on LLaMA. Acc (\%) is the accuracy of the downstream task.}
		\label{tab:llama_quant}
		\scalebox{0.8}{\begin{tabular}{ccc}
				\hline
				Quantization     & Acc & WR \\ \hline
				4-bit & 95.996 & 100.0 \\
				8-bit & 96.148 & 100.0 \\ \hline
		\end{tabular}}

	\end{minipage}
	
\end{figure}

\subsubsection{White-box Adversaries}
\label{subsec:white-box}

A white-box adversary can be encountered in the use case where the model backbone parameters are leaked or made public. We take into consideration the following white-box attacks: fine-tuning, pruning, and quantization, consistent with the previous watermarking literature.

\textbf{Fine-tuning}. In previous watermarking works \cite{lukas2022sok}, fine-tuning only employs minimal learning rates and few iterations. To make the scene resemble an actual attack, we reuse the setup in Sec.~\ref{subsec:fidelity} where the fine-tuning takes 5 epochs ensuring the downstream tasks are truly optimized. For a more objective view on the robustness against fine-tuning attack, we record the watermark rates on the extraction sets for every fine-tuning epoch.

The WRs under fine-tuning attacks on different CV models are presented in the leftmost three column of Fig.~\ref{fig:white-box} and that on different NLP models in Table~\ref{tab:finetune-NLP}. Across all settings, TokenMark-B and TokenMark-S achieve almost 100\% WRs throughout the fine-tuning process. The CLIP model has high robustness for all methods, probably because the pre-trained CLIP fuses texts which stick to outliers (backdoors). SSLGuard fails to maintain the watermark performance on Cifar10 and CelebA as fine-tuning proceeds, showing its lack of robustness against the attack.

We also compare the watermark robustness between CNN and ViT. As shown in Fig.~\ref{fig:cnnvsvit}, our observation is consistent with previously reported results, indicating that it is easier to embed and remove a watermark in/from Transformer-based models. 

We calculate the averages of 3 pre-trained models and show the overall robustness in Fig.~\ref{fig:thumbnail}.

\textbf{Pruning}. It is claimed in \cite{lukas2022sok} that if partial neurons are disabled by pruning, the watermark could be removed from the model. A pruning ratio $r$ is set, and $r$ of the neurons in each layer are pruned. We set $r$ ranging from 0.1 to 0.5 with a step of 0.1.

Results are shown in the fourth column of Fig.~\ref{fig:white-box}. It is observed that TokenMark-S is highly robust against pruning, and TokenMark-B only fails Dino V2, but still is better than trigger-based baseline. Trigger-based watermarking shows great unstability under pruning. Note that the fall-and-rise pattern of SSLGuard on CLIP is also observed in the contrastive CNN version of SSLGuard in their experiments \cite{cong2022sslguard}. Pruning on LLaMA 2 7B does not affect the WR of TokenMark-B much as shown in Fig.~\ref{fig:llama_prune}, considering the accuracy decay to random guess on the downstream task.

\textbf{Quantization} compresses the model by changing the weights to a lower bit representation, which could be leveraged by the adversary to remove the watermark. We reuse the implementation of \cite{lukas2022sok} for quantization attack. We consider the downstream (fine-tuning) accuracies and the watermark rates under 5-bit to 1-bit quantization. The results are reported in the rightmost column of Fig.~\ref{fig:white-box}. TokenMark remains robust against severe quantization, while backdoor-based watermark, especially SSLGuard, shows an unpredictable rise-and-fall pattern, casting doubt on its robustness against model quantization. For large language model LLaMA 2 7B, we leverage the LLM quantization tool, bitsandbytes\footnote{https://github.com/bitsandbytes-foundation/bitsandbytes}, to quantize LLaMA 2 7B into 4-bit and 8-bit. The watermark remains robust, as shown in Table~\ref{tab:llama_quant}.

Note that TokenMark may seem vulnerable in LLM given that an adversary with the embedding vocabulary can search the permutation matrix by brute-force. However, the adoption of the absolute position embedding to the input tokens prevents such an attack due to unknown embedding weights.

\subsubsection{Black-box Adversaries}
\label{subsec:black-box}
As reported in \cite{lukas2022sok}, watermarks tend to vanish under model extraction attack, since the extracted model might not carry any watermarking functionality. We adopt the same extraction attack as in \cite{cong2022sslguard}, using CelebA as the adversary's extracting dataset, to evaluate the robustness of TokenMark. The extraction attacker aligns the feature of its extracted model with that of the watermarked model, by minimizing the cosine similarity between the model outputs. SGD optimizers with learning rate $10^{-4}$ for ViT-timm and $10^{-5}$ for ViT-Dino v2 and ViT-CLIP are applied for optimization. 

As shown in Table~\ref{tab:extracting}, on the extracted models, the WR remains as high as 100\%, showing the watermarks are retained after extraction. This can be attributed to the fact that the features of the watermarked models not only carry the main functionality of the pre-trained model but also the watermarking functionality. It shows the strength of TokenMark in deeply associating the watermark with the feature representation of pre-trained models, making it almost impossible for a black-box attacker to steal the model without removing or bypassing the watermark.

\begin{table}[]
	\caption{Performance under black-box extraction attack: downstream-task Acc (\%) and watermark rates (\%) on extracted ViTs.  }
	\label{tab:extracting}
	\centering
	\scalebox{1.0}{\begin{tabular}{cccc}\hline
			Acc/WR& ViT-timm & ViT-Dino v2 & ViT-CLIP \\\hline
			TokenMark-B & 95.8/100.0    & 91.7/100.0        & 86.7/100.0  \\
			TokenMark-S & 95.5/100.0    & 94.6/100.0        & 89.2/100.0    \\\hline
	\end{tabular}}
\end{table}

\subsubsection{Adaptive Adversaries}
\label{subsec:adaptive}
Finally, we assume an \textbf{adaptive adversary}, who has some knowledge of TokenMark-S, including $\theta_*$, $sk$, $G(\cdot)$, except for $\mP$. Leveraging this knowledge, the adversary could try to overwrite the existing watermark, or to remove the watermark by reverse-engineering the permutation matrix. We would like to note that this adaptive setting is unrealistic for the attacker, as both $sk$ and $G(\cdot)$ are assumed to be kept secret by the authority. However, we still include this imaginary adversary to evaluate the robustness of TokenMark.

\textbf{Overwriting}. We first assume the adversary tries to embed another watermark with the same strategy, in an attempt to overwrite the original watermark. In TokenMark, the adversary overwrites the watermark by fine-tuning the model with another permutation matrix $\mP'$. The overwriting inherits the same configuration with that of the original embedding process, including the hyperparameters and fine-tuning steps. We evaluate the WRs of the original watermark. 

Results in Table~\ref{tab:overwrite} show that the overwriting fails to eliminate the original watermark. Hence TokenMark is robust to overwriting attacks. We analyze that this is mostly due to the redundant capacity in Transformer-based models to incorporate multiple sets of parameters in $\Theta_P$ of Eq.~(\ref{eq:effect}). We believe multiple permutation matrices could co-exist within one model.

The robustness to overwriting attacks can be further extended to the resilience against the ambiguity attack. In resolving the dispute of ownership, the model with only one watermark extracted is the original one, and the watermark belongs to the owner. 

\begin{table}[h]
	\caption{Performance against overwriting attack: watermark rates (\%) on pre-trained models overwritten by Cifar10. The embedding and extraction sets are Cifar10.}
	\label{tab:overwrite}
	\centering
	\scalebox{1.0}{
		\begin{tabular}{cccc}
			\hline
			WR & ViT-timm & ViT-Dino v2 & ViT-CLIP \\\hline
			TokenMark-B & 100.0      & 100.0    & 100.0      \\
			TokenMark-S & 100.0	& 100.0& 100.0    \\\hline
	\end{tabular}}
\end{table}

\textbf{Adaptive Removal.} We further assume the adversary tries to remove the watermark by first finding a surrogate permutation matrix $\mP'$ and then use $\mP'$ to facilitate a watermark removal attack. To find such a $\mP'$, we assume the adversary could perform random search or gradient-based search.

\textit{1) Random search.} We allow the attacker to randomly generate a batch of $\mP'$s as the permutation matrices, in an effort to extract the watermark from $\theta_*$. If the adversary obtains a WR greater than 0.5 on a very small dataset, it then verifies that permutation on the entire extraction set. If the WR is still greater than 0.5, we consider the random search successful.

The adversary adopted the CelebA and Cifar10 datasets, and tried out about 200,000 randomly generated permutation matrices, but none of them can be decoded. Most of the decoded vectors are orthogonal to $sk$. Due to the failure of finding a valid $\mP'$, the adversary could not perform the subsequent removal attack either. This can be explained by the massive permutation space (see the discussion in Sec.~\ref{sec:discussion-on-permutation}).

% It can be further concluded that $\theta_*$, $sk$, $G(\cdot)$ leak no information about the permutation matrix. The results indicate that TokenMark is robust against adaptive attacks. 

% \blue{\textit{Gradient-based search.} We use a randomly initialized parameter $\mP'$ with the same size as $\mP$, and optimize it with $\Ls_{atk}=-\cos(G(F(Z, P'(\theta_*)),sk))+\mathrm{MSE}(\mI, \mP'^\top\mP')$ on Cifar10 dataset. Further, we constrain each row of $\mP'$ to be a unit vector by picking the one with the largest value as 1 and the rest as 0 after each Adam optimizer step, in order to make sure that $\mP'$ is a permutation matrix. Learning rates from $10^{-7}$ to $10^1$ are tried. The result shows that $\mathrm{MSE}(\mI, \mP'^\top\mP')$ converges to $0$ in no time but the cosine similarity stuck at $-0.1$. This result is expected since the optimization problem here is discrete.}

\textit{2) Gradient-based search and adaptive removal.} We further consider an adversary who first finds a surrogate permutation matrix $\mP'$ via gradient-based optimization, and then uses $\mP'$ to instantiate a watermark removal attack. The adversary first initializes and obtains a matrix $\mP'$ by minimizing
\begin{equation*}
	\mathcal{L}_{a} = -\cos(G(F(\mZ, P'(\theta_*))),sk) + \alpha\|\mI - \sigma(\mP'^\top)\sigma(\mP')\|_2^2,
\end{equation*}
where $\sigma(\cdot)$ denotes the Softmax activation function and $\alpha=0.1$ is a hyper-parameter. We ensure each row of $\mP'$ to be a one-hot vector via Gumbel-Softmax sampling~\cite{jang2016categorical}, which is a commonly used trick for sampling discrete one-hot vectors while allowing for gradient back-propagation~\cite{abdelnabi2021awt}. We run the optimization on Cifar10 dataset for 3 epochs, and convert the obtained $\mP'$ into a 0-1 matrix by setting the largest value of each row as 1 and the rest as 0. Note that the obtained $\bm{P}'$ is not necessarily a permutation matrix, but thanks to the $L_2$ regularization in $\mathcal{L}_a$, it is \emph{almost} one, with only a few duplicated or missing rows. The adversary could easily make $\bm{P}'$ a valid permutation matrix by randomly replacing duplicated rows with missing elements at a negligible loss.

The adversary could then use its obtained $\mP'$ to facilitate a watermark removal attack, by training a de-watermarked model parametrized by $\hat{\theta}$ to minimize the loss
\begin{equation*}
	\mathcal{L}_{rm} = -\mathrm{sim}(F(\mZ, \hat{\theta}), F(\mZ, \theta_*))
	+ \textrm{sim}^2(sk, G(F(\mZ, P'(\hat{\theta})))),
\end{equation*}
where $\hat{\theta}$ is initialized from the watermarked model $\theta_*$.
\iffalse
\begin{table}[htb]
	\caption{Performance against gradient-based search and adaptive removal attack: watermark rates on original and de-watermarked ViT-timm, using both the original $\mP$ and the adversary's $\mP'$.}
	\label{tab:adaptive-removal-attack}
	\centering
	\begin{tabular}{ccc}
		\hline
		& WR ($\mP$) & WR ($\mP'$) \\ \hline
		Original       & 1.0        & 1.0              \\
		Removal Atk.   & 0.9998     & 0.0              \\ \hline
	\end{tabular}
\end{table}
\fi

We perform gradient-based search on ViT-timm and find that the adversary could indeed find a permutation matrix $\mP'$ that achieves $1.0$ watermark rate. However, this $\mP'$ differs from $\mP$ significantly, with only $0.13\%$ identical elements. As a result of this difference, the removal attempt is not successful: the watermark rate on the de-watermarked model $\hat{\theta}$ w.r.t. $\mP'$ is reduced to $0.0$, but the rate w.r.t. the original $\mP$ is still almost $1.0$. The adversary ends up removing its own $\mP'$ but not the original watermark. Additionally, since the real permutation $\mP$ is also stored at the authority side, it is also impossible for the adversary to cast ambiguity on the model ownership, as $\mP'$ differs from $\mP$ and will be immediately identified by the authority as a fake permutation matrix.

\subsubsection{Discussion on Robustness Results} 
We now provide a deeper analysis on the reason behind TokenMark's robustness. 
First, as has been stated in Sec.~\ref{sec:discussion-on-permutation}, TokenMark's watermark is structure-driven --- embedding the watermark by the permutation order of model weights, which are barely associated with a particular dataset. On the other hand, trigger-based schemes, represented by Adi et al.\cite{adi2018turning} and SSLGuard\cite{cong2022sslguard}, are data-driven --- relying heavily on the trigger patterns. Hence the model weights are tied to the trigger set. Once the model parameters are pruned or fine-tuned, some `key' neurons may be modified and thus the triggering path disappears. However, such changes hardly affect the permutation order of the weights, thus preserving TokenMark's watermarking functionality. Second, Transformer-based models are more vulnerable to backdoors \cite{doan2023defending,yuan2023you}, meaning that it is easier to implant or remove a backdoor for Transformer models. Transformers are thereby less robust in trigger-based watermarking, compared to the ResNet models used in the original SSLGuard. Third, in face of black-box model extraction attack, the trigger-based scheme hinges upon the trigger data distribution which could be quite different from the extracting data used by the attacker, and hence the extracted model easily removes the watermark without affecting the major functionality of the model. In contrast, the watermark of TokenMark is closely tied up with the weights, and the extracted model would largely preserve the watermark as an inherent feature.

We provide an intuition on why it is hard for the adversary to invert $\mP$. Assuming the backbone $F$ and decoder $G$ only contain linear layers, the adversary essentially solves the following equations to obtain $\mP'$:
$$G\cdot F(\mZ, P'(\theta_*)) = sk. $$
The unknowns of the equations are collectively represented by $\mP'$ which is orders of magnitude larger than the dimension of $sk$ (256 in TokenMark-S and 10 in TokenMark-B). Hence the equations would have many solutions, and it is almost impossible for the adversary to identify the $\mP$.

\subsection{Study of Hyperparameters}
\label{subsec:ablation}
We study how the change of some hyperparameters affects TokenMark's performance.

\begin{figure}
	\centering
	\begin{subfigure}[b]{0.49\linewidth}
		\includegraphics[width=\linewidth]{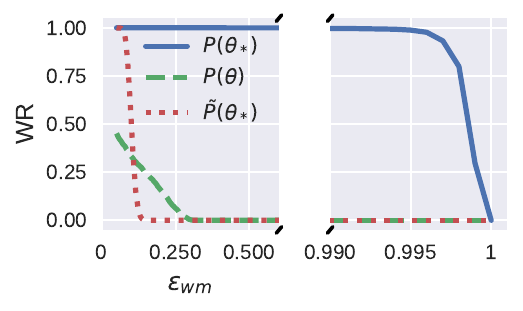}
		\caption{ViT-timm}
		\label{fig:threshold_ViT_timm}
	\end{subfigure}
	\begin{subfigure}[b]{0.49\linewidth}
		\includegraphics[width=\linewidth]{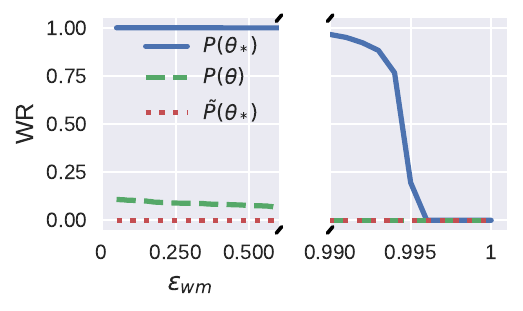}
		\caption{ViT-Dino v2}
		\label{fig:threshold_ViT_Dinov2}
	\end{subfigure}
	\caption{Watermark rates with different $\epsilon_{wm}$s in TokenMark-S.  $P(\theta_*), P(\theta), \tilde{P}(\theta_*)$ represent the watermarked model, unwatermarked model, and the incorrectly decoded model, respectively. }
	\label{fig:threshold}
\end{figure}

\textbf{Watermark threshold} $\epsilon_{wm}$ is set to 0.5 by default in measuring WR. Here we vary the value from 0.050 to 0.999 in extracting the watermark from ViT-timm, ViT-Dinov2 by Cifar10, both using the watermarked model ($P(\theta_*)$), the unwatermarked model ($P(\theta)$) and the incorrectly extracted one ($\tilde{P}(\theta_*)$) in TokenMark-S. As shown in Fig.~\ref{fig:threshold}, for ViT-timm, the threshold does not affect WRs where $0.30 < \epsilon_{wm} < 0.997$. The safe region for Dinov2 is $\epsilon_{wm} < 0.995$. A too-low threshold incurs false positives on both unwatermarked and incorrectly decoded model while a too high value leads to false negatives. Overall, most decoded vectors either have a very high similarity, or almost zero similarity to the secret vector. Hence TokenMark is quite robust to the change in $\epsilon_{wm}$.

\textbf{Size of extraction set} also has little influence on the watermark rate. To observe its influence, we specifically select watermark extraction from ViT-timm in TokenMark-S with $\epsilon_{wm} = 0.998$ at which the WR is shy of 1.00. In that case, we vary the Cifar10 extraction set size from 10 to 10000. The change of WR is mostly observed to follow the law of large numbers and it converges to 0.52.

\begin{figure}[h]
	\centering
	\begin{subfigure}[b]{0.49\linewidth}
		\includegraphics[width=\linewidth]{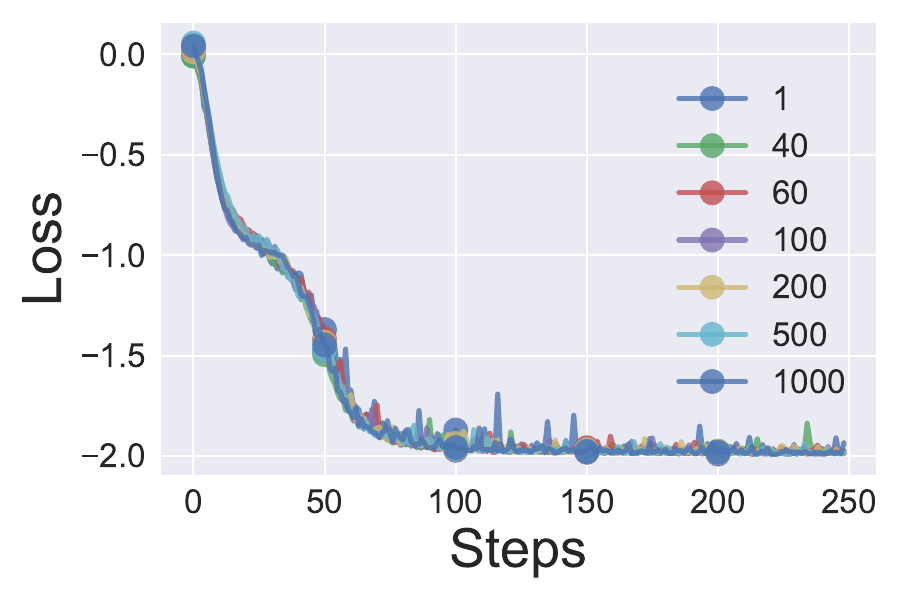}
		\caption{Curve of Backbone}
		\label{fig:loss_B}
	\end{subfigure}
	\begin{subfigure}[b]{0.49\linewidth}
		\includegraphics[width=\linewidth]{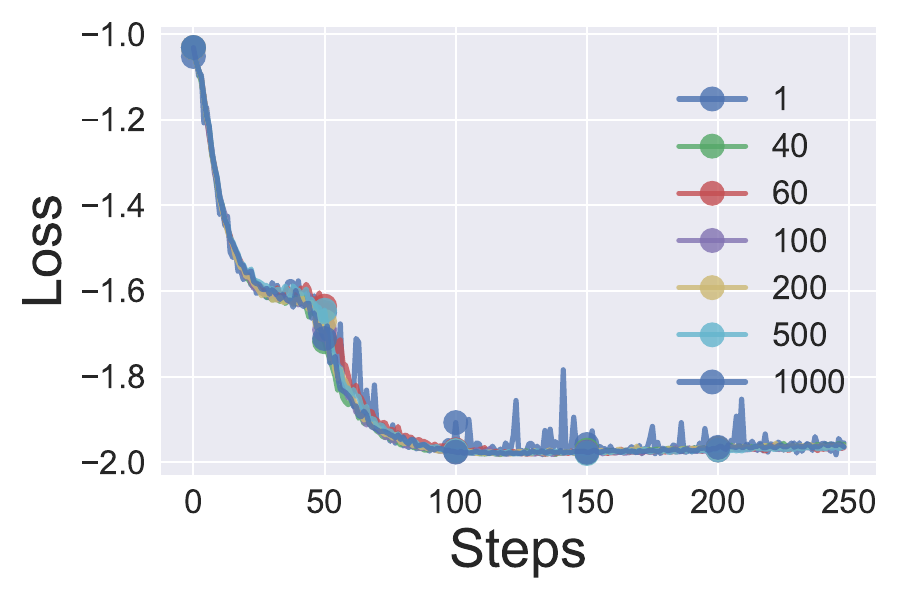}
		\caption{Curve of Watermark head}
		\label{fig:loss_G}
	\end{subfigure}
	\caption{Loss curves of embedding TokenMark-S watermark with different embedding set sizes.}
	\label{fig:embedding_size}
\end{figure}

\textbf{Size of embedding set}. We have tried embedding TokenMark-S watermarks using fewer Cifar10 images (1000, 500, 200, 100, 60, 40 and 1) and more training epochs (5, 10, 25, 50, 85, 125, and 250) accordingly. The embedding loss curves are not much different from one another, as shown in Fig.~\ref{fig:embedding_size}. Its effectiveness (a) and fidelity performance (b) are as good as the default setting, but the robustness decays given too few embedding samples. Taking single-image embedding as an example, its watermark survived fine-tuning on Cifar10 and STL10, but not CelebA, nor pruning attack ($r > 0.2$). A possible explanation is that too few data records may induce memorization of outlier examples, and in this situation, the watermark may degenerate into a data-driven one violating the principle of TokenMark.

% \section{Limitations and Future Work}

% The limitation of TokenMark is apparent, in that it only suits Transformer-based models, excluding convolutionaloperation. The same properties cannot be obserbed when the equivariance is break by any means, like

\section{Conclusion}
\label{sec:conclusion}

We proposed TokenMark, a watermarking scheme for pre-trained models, which is independent of the data modality, the downstream task, and the trigger set. The design is inspired by the permutation equivariance property of Transformers, and could be adopted to enhance a variety of backdoor-based watermarking systems in terms of robustness and efficiency. Experimental results on a range of pre-trained models and datasets have shown that TokenMark's superiority over the state-of-the-art approaches.

% \section*{Acknowledgments}
% The research was supported in part by National Science and Technology Major Project 2021ZD0112801, NSF China (62272306, 62032020, 62136006).

% {\appendix[Proof of the Zonklar Equations]
% Use $\backslash${\tt{appendix}} if you have a single appendix:
% Do not use $\backslash${\tt{section}} anymore after $\backslash${\tt{appendix}}, only $\backslash${\tt{section*}}.
% If you have multiple appendixes use $\backslash${\tt{appendices}} then use $\backslash${\tt{section}} to start each appendix.
% You must declare a $\backslash${\tt{section}} before using any $\backslash${\tt{subsection}} or using $\backslash${\tt{label}} ($\backslash${\tt{appendices}} by itself
%  starts a section numbered zero.)}

%{\appendices
%\section*{Proof of the First Zonklar Equation}
%Appendix one text goes here.
% You can choose not to have a title for an appendix if you want by leaving the argument blank
%\section*{Proof of the Second Zonklar Equation}
%Appendix two text goes here.}

% \section{References Section}

\bibliography{bodyguard}
\bibliographystyle{IEEEtran}

 % argument is your BibTeX string definitions and bibliography database(s)
%\bibliography{IEEEabrv,../bib/paper}
%

% If you have an EPS/PDF photo (graphicx package needed), extra braces are
%  needed around the contents of the optional argument to biography to prevent
%  the LaTeX parser from getting confused when it sees the complicated
%  $\backslash${\tt{includegraphics}} command within an optional argument. (You can create
%  your own custom macro containing the $\backslash${\tt{includegraphics}} command to make things
%  simpler here.)
 
% \vspace{11pt}

% \bf{If you include a photo:}\vspace{-33pt}
% \begin{IEEEbiography}[{\includegraphics[width=1in,height=1.25in,clip,keepaspectratio]{fig1}}]{Michael Shell}
% Use $\backslash${\tt{begin\{IEEEbiography\}}} and then for the 1st argument use $\backslash${\tt{includegraphics}} to declare and link the author photo.
% Use the author name as the 3rd argument followed by the biography text.
% \end{IEEEbiography}

% \vspace{11pt}

% \bf{If you will not include a photo:}\vspace{-33pt}
% \begin{IEEEbiographynophoto}{John Doe}
% Use $\backslash${\tt{begin\{IEEEbiographynophoto\}}} and the author name as the argument followed by the biography text.
% \end{IEEEbiographynophoto}

% \newpage

\appendix
% \clearpage
\setcounter{page}{1}

\section{Details of TokenMark-S}
\label{sec:app_alg}

\subsection{The Embedding}

The embedding process is a fine-tuning process. Before the embedding, the owner selects a secret permutation matrix $\mP$ and a secret vector $sk$. The embedding dataset can be any valid dataset. In each iteration of fine-tuning, a batch of embedding data is sampled to perform three-step updates sequentially: 1) minimizing $\Ls_{S}$ over $\theta_s$; 2) minimizing $\Ls_{G}$ over $G$; 3) minimizing $\Ls_{B}$ over $\theta_*$. After the embedding, the permutation matrix $\mP$, the secret vector $sk$, and the decoder $G(\cdot)$ are kept secret and submitted to the authority, while the watermarked model $\theta_*$ or its API access is published. Algorithm~\ref{alg:embedding} provides more detail.

\begin{algorithm}
    %\textsl{}\setstretch{1.8}
    \renewcommand{\algorithmicrequire}{\textbf{Input:}}
    \renewcommand{\algorithmicensure}{\textbf{Output:}}
    \caption{Watermark Embedding}
    \label{alg:embedding}
    \begin{algorithmic}[1]
        \REQUIRE 1) The original pre-trained model $\theta$; 2) randomly-generated (or specifically-designed) permutation matrix $\mP$ and secret vector $sk$; 3) embedding dataset $B$. 
        \ENSURE 1) Decoder $G$; 2) watermarked model $\theta_*$.
        \STATE Initialize $\theta_*$ with $\theta$. Initialize $\theta_s$ and $G$ randomly.
        \REPEAT
        \STATE Sample a batch of data $\mX$ from $B$ and get input embeddings $\mZ = T(\mX)$
        \STATE Update $\theta_s$ w.r.t. $\Ls_S$ for one step
        \STATE Update $G$ w.r.t. $\Ls_G$ for one step
        \STATE Update $\theta_*$ w.r.t. $\Ls_B$ for one step
        \UNTIL Losses converge.
    \end{algorithmic}  
\end{algorithm}

\subsection{The Extraction}
The watermark extraction can be done with white/black-box access to the watermarked model, and API access would suffice for the verification. The extraction set can be any valid dataset, not necessarily the same with the embedding set. The prover only needs to permute its input feature with $\mP$ and send it to the target model for inference. Applying the forward permutation equivariance in Eq.~(\ref{eq:forward_equivariance}), the inference process is equivalent to inferring $\mZ$ on $P(\theta_*)$. Then output feature is restored by $\mP^{-1}$ and is fed into $G$ to decode. The cosine similarity between the decoded vector and the secret vector is calculated. If the similarity is higher than a threshold $\epsilon_{wm}$, the watermark is considered as detected. By calculating the percentage of watermark detected (watermarking rate) in the extraction dataset, the authority can determine whether the pre-trained model belongs to the owner. The detail of the extraction is shown in Algorithm~\ref{alg:extraction}.

\begin{algorithm}
    %\textsl{}\setstretch{1.8}
    \renewcommand{\algorithmicrequire}{\textbf{Input:}}
    \renewcommand{\algorithmicensure}{\textbf{Output:}}
    \caption{Watermark Extraction}
    \label{alg:extraction}
    \begin{algorithmic}[1]
        \REQUIRE 1) Suspected model $\theta'_*$; 2) permutation matrix $\mP$, secret vector $sk$, and decoder $G$; 3) extraction dataset $D$. 
        \ENSURE Watermarking rate WR.
        \STATE count = 0, total = 0
        \REPEAT
        \STATE Pick and remove a data record $x$ from $D$ and get input embedding $\mZ = T(x)$ 
        \IF {$\textrm{sim}(G(F(\mP\mZ, \theta'_*)\mP^{-1}), sk) > \epsilon_{wm}$}
            \STATE count += 1
        \ENDIF
        \STATE total += 1
        \UNTIL $D = \emptyset$
        \STATE WR = count / total.
    \end{algorithmic}  
\end{algorithm}

\section{Experimental Details}
\label{sec:app_exp}

\subsection{Supplementary Setup}
\label{sec:app_setup}

% To be re-calculated. huge project qwq
% done
\begin{table*}[t]
    \caption{Performance of downstream CV tasks on watermarked pre-trained models. Accuracy (\%) is used as the metric. In bracket is the accuracy gap compared to the original models (original minus watermarked).  }
    \label{tab:fidelity}
    \centering
    \scalebox{0.6}{\begin{tabular}{ccccccccccccc}
        \hline
        Downstream & \multicolumn{3}{c}{Cifar10} & \multicolumn{3}{c}{STL10} & \multicolumn{3}{c}{CelebA}                              & \multicolumn{3}{c}{Tiny-ImageNet}                                             \\ 
        Model        & timm    & Dino V2    & CLIP  & timm    & Dino V2    & CLIP    & timm    & Dino V2    & CLIP & timm    & Dino V2    & CLIP\\ \hline
        Adi et al. & 98.69(-1.13) & 98.67(0.03) & 97.48(0.23)	& 99.05(-1.00) & 99.43(-0.65) & 97.15(+0.06)	& 91.23(+0.75) & 92.05(-0.77) & 92.07(-0.04)	& 90.87(-0.29) & 89.11(0.1) & 80.83(-1.32) \\
        TokenMark-B     & 98.05(-0.49) & 98.47(+0.23) & 98.13(-0.42)	& 98.61(-0.56) & 98.53(0.75) & 97.54(-0.33)	& 92.01(-0.03) & 92.20(-0.92) & 92.06(-0.03)	& 90.89(-0.31) & 88.94(+0.34) & 79.09(+0.42) \\ 
        SSLGuard  & 98.04(-0.35)  & 99.11(-0.41) & 97.48(+0.23)	& 98.69(-0.70) & 99.19(+0.09) & 97.44(-0.23)	& 91.92(+0.06) & 92.27(-0.91) & 91.96(+0.05)	& 91.24(-0.66) & 89.15(+0.23) & 80.57(-1.44)\\ 
        TokenMark-S    & 97.47(+0.18) & 98.82(-0.12) & 97.65(+0.06)	& 98.02(+0.03) & 99.25(+0.03) & 97.45(-0.24)	& 91.82(+0.16) & 92.18(+0.10)  & 92.06(-0.03)	& 90.40(+0.18)  & 88.99(+0.29)  & 79.86(-0.35)\\
        \hline
    \end{tabular}}
\end{table*}

\begin{table*}[t]
    \caption{Performance of downstream NLP tasks on watermarked pre-trained models. Metrics of IMDB and SQuAD are accuracy (\%). The metric of SWAG is Exact Match (EM) (\%) while that of WikiText2 is perplexity. The former three metrics are the higher the better, while the latter is the lower the better. In bracket is the performance gap compared to the original models (original minus watermarked). }
    \label{tab:fidelity_NLP}
    \centering
    \scalebox{0.8}{\begin{tabular}{ccccccc}
        \hline
        Downstream & \multicolumn{2}{l}{IMDB} & \multicolumn{2}{l}{SQuAD}                              & SWAG & WikiText2                                        \\ 
        Model        & BERT    & GPT-2    & BERT    & GPT-2 & BERT    & GPT-2\\ \hline
        TokenMark-B    & 94.00(-0.13)   & 93.85(-0.37)  & 80.63(-0.81)  & 67.71(-0.06)  & 80.47(-0.15)  & 30.20(+0.00) \\
        TokenMark-S   & 93.95(-0.08)   & 93.58(-0.10)  &  79.89(-0.07)  &  67.89(-0.24)  & 80.10(+0.22)  &  30.14(-0.06)   \\
        \hline 
    \end{tabular}}
\end{table*}

\textbf{Datasets and downstream tasks}. 

\textit{Cifar10} \cite{krizhevsky2009learning} contains 60,000 colored images of size $3\times32\times32$ from 10 categories, with 6,000 images per class. Adam optimizer is used in fine-tuning with a learning rate of $10^{-4}$ for ViT-timm and $10^{-5}$ for the rest.  
%The fine-tuning task is 10-class classification and accuracy is used as the evaluation metric. The downstream head used for Cifar10 is a 768 to 10 linear layer.

%\textit{ImageNet} \cite{deng2009imagenet} is a large-scale dataset for object detection, localization, and classification, containing 1.2 million images in 1,000 classes. We use this dataset for watermark extraction.

\textit{Tiny-ImageNet} is a subset of ImageNet, containing 200 classes, 500 training images, and 50 validation images per class. An Adam optimizer is used in fine-tuning with a learning rate $10^{-5}$ for all vision models. %\textit{It is worth noting that the timm pre-trained models are pre-trained on ImageNet, which brings it some advantages when fine-tuning on this dataset. This can be viewed as the scenario where the attacker has access to some of the pre-training data.}

\textit{STL10} \cite{coates2011analysis} is a subset of ImageNet, containing 10 classes, 5,000 training images and 8,000 test images. Each image is of size $3\times96\times96$. An Adam optimizer is used in fine-tuning with a learning rate $10^{-4}$ for ViT-timm and $10^{-5}$ for the rest.  %\textit{Worth noting that the timm pre-trained models are pre-trained on ImageNet, which brings it some advantages when fine-tuning on this dataset. This can be viewed as the scenario where the attacker has access to some of the pre-training data.}

%This dataset is used for 10-class classification fine-tuning. Accuracy is used as the evaluation metric. The downstream head used for STL10 is a 768 to 10 linear layer.

\textit{CelebA} \cite{liu2015deep} is a large-scale face attributes dataset with more than 200K celebrity face images, each with 40 attribute annotations. An Adam optimizer is used in fine-tuning with a learning rate $10^{-4}$ for ViT-timm and $10^{-5}$ for the rest. % We use this dataset for attribute classification fine-tuning. Accuracy is used as the evaluation metric. The downstream head used for CelebA is a 768 to 40 linear layer.

\textit{QNLI} \cite{wang2018glue} is a Natural Language Inference dataset automatically derived from the Stanford Question Answering Dataset v1.1 (SQuAD) and is part of the GLUE benchmark. % We use the corpus from this dataset for watermarking the NLP model. 

\textit{IMDB} \cite{IMDB} is a dataset for binary sentiment classification. It consists of 50,000 movie reviews from IMDB, with 25,000 for training and 25,000 for testing.  An AdamW optimizer with a learning rate $2 \times 10^{-5}$ is adopted.% We use this dataset for binary classification fine-tuning. Accuracy is used as the evaluation metric. The downstream head used for IMDB is Huggingface's sequence classification head of Huggingface Transformer.

\textit{SQuAD} \cite{rajpurkar2016squad} is a reading comprehension dataset, consisting of questions posed by crowdworkers on a collection of Wikipedia articles, where the answer to every question is a segment of text, or span, from the corresponding reading passage. An AdamW optimizer with a learning rate $2 \times 10^{-5}$ is adopted.% We use this dataset for question and answering fine-tuning. Exact match (EM) and F1 score are used as the evaluation metrics. The downstream head used for SQuAD is a question answering head of the Huggingface Transformer.

\textit{SWAG} \cite{zellers2018swag} is a dataset for grounded commonsense inference, unifying natural language inference and physically grounded reasoning, which contains 113k multiple-choice questions about grounded situations.  An AdamW optimizer with a learning rate $5 \times 10^{-5}$ is adopted for training.% We use this dataset for BERT multiple-choice fine-tuning. Accuracy is used as the evaluation metric. The downstream head used for SWAG is the multiple-choice head of the Huggingface Transformer.

\textit{WikiText2} \cite{merity2016pointer} is a language modeling dataset, which is a collection of over 100 million tokens extracted from the set of verified Good and Featured articles on Wikipedia. An AdamW optimizer with a learning rate $5 \times 10^{-5}$ is adopted for training.% We use this dataset for GPT-2 generation fine-tuning. Perplexity is used as the evaluation metric. The downstream head used for WikiText2 original generative language modeling head of GPT-2.

\textbf{Model architecture and watermark hyperparameters}. All the Transformers in this work is base-sized, i.e. 12 layers of 768 dimension Transformer blocks with 12 head and 3072 hidden dimension.

All the downstream heads are a randomly initialized linear layer from 768 to number of classes. The watermarking hyperparameters for TokenMark-S and SSLGuard are as follows: the secret vector $sk$ is a 256-dimensional vector randomly generated from a standard normal distribution. The watermark decoder $G$ is a 2-layer Multi-Layer Perceptron (MLP) with a hidden layer of size 256 and ReLU activation. The first token of the output of the transformer backbone is used for decoding the watermark. Additionally, dropout mechanism is applied on the output of each Transformer block in the embedding process to enhance the robustness of watermarking.

Trigger pattern for Adi et al. is a pure white patch of size $16\times16$ at the top-left corner of the $224\times224$ image, which is designed according to \cite{yuan2023you}. And the trigger pattern for SSLGuard is from its original work \cite{cong2022sslguard}.

\textbf{Setup of Fig.~\ref{fig:thumbnail}}. Actually, there is no additional experiments conducted for Fig.~\ref{fig:thumbnail}. The middle subfigure is the averaged results of  experiment on robustness. The results of fine-tuning is averages of the final epoch WR of 3 models, and the results of pruning and quantization are the averages of all the data point, including the WR of all the pruning ratios/quantization bits of 3 models. The rightmost subfigure is the recorded time consumption of ViT-timm.

\textbf{Setup of Fig.~\ref{fig:cnnvsvit}}. The embedding setup for Fig.~\ref{fig:cnnvsvit} is different from the setup in main experiments. A poisoning rate of 0.5, which is leveraged in the main setup, would cause severe ultility degradation of CNN. We set the poisoning rate to 0.1 and extend the poisoned training to 5 epochs. The CNN architecture we chose is ResNet-50, which has a similar number of parameters with ViT-base. Pre-trained model is provided by timm. An adam optimizer with a learning rate of $10^{-4}$ is used for poisoned fine-tuning.

\subsection{Results on Watermark Fidelity}
\label{sec:app_fidelity}

The supplementary results are shown in Table~\ref{tab:fidelity} (CV) and Table~\ref{tab:fidelity_NLP} (NLP). Metrics for CV tasks, IMDB and SQuaD are accuracy. The metric for SWAG is exact match, and that for WikiText2 is perplexity. The perplexity is the lower the better, while all other metrics are the higher the better. Detailed fine-tuning hyperparameters are provided in Sec.~\ref{sec:app_setup}.

\vfill

\end{document}